\documentclass[preprint,12pt]{elsarticle}

\usepackage{amssymb}
\usepackage{amsmath,amsfonts}
\usepackage{algorithmic}
\usepackage{algorithm}
\usepackage{array}
\usepackage[caption=false,font=normalsize,labelfont=sf,textfont=sf]{subfig}
\usepackage{textcomp}
\usepackage{stfloats}
\usepackage{url}
\usepackage{verbatim}
\usepackage{graphicx}
\usepackage{cite}
\usepackage{color}
\date{March 22 2022}

\journal{Automatica}

\begin{document}

\begin{frontmatter}



\title{Adaptive Unscented Kalman Filter under Minimum Error Entropy with Fiducial Points \\ for Non-Gaussian Systems(\textit{This work has been submitted to the Automatica on 22 March
2022 for possible publication. Copyright may be transferred
without notice, after which this version may no longer be
accessible})}


\author[inst1]{Boyu Tian}

\affiliation[inst1]{organization={Key Laboratory of Magnetic Suspension Technology and Maglev Vehicle, Ministry of Education},
            addressline={School of Electrical Engineering Southwest Jiaotong University}, 
            city={Chengdu},
            country={China}}

\author[inst1]{Haiquan Zhao  \corref{cor1}}

\cortext[cor1]{Corresponding author\\ E-mail addresses:: bytian\_swjtu@126.com(B.Y. Tian), hqzhao\_swjtu@126.com(H.Q. Zhao)}

\begin{abstract}
The minimum error entropy (MEE) has been extensively used in unscented
Kalman filter (UKF) to handle \textcolor{black}{impulsive noises or abnormal measurement
data} in non-Gaussian systems. However, the
MEE-UKF has poor numerical stability due to the inverse operation of singular matrix. In this paper, a novel UKF based on minimum error entropy with fiducial  points (MEEF) is proposed \textcolor{black}{to improve the problem of non-positive
definite key matrix.  By adding the correntropy to the error entropy, the proposed algorithm further enhances the ability of suppressing impulse noise and outliers.}
 \textcolor{black}{At the same time, }
 considering the uncertainty of noise distribution, the modified Sage-Husa estimator of noise statistics is introduced to adaptively update the noise covariance matrix. In addition, the convergence analysis of the proposed algorithm provides a guidance for the selection of kernel width. The robustness and estimation accuracy of the proposed algorithm are manifested by the state tracking examples under  complex non-Gaussian noises.
\end{abstract}




\end{frontmatter}


\section{Introduction}

Correntropy, as  similarity measure of two random variables \citep{liu2007correntropy}, \textcolor{black}{has been widely used in various fields}, such as  machine learning \citep{chen2018mixture,chen2021multikernel}, adaptive filtering \citep{chen2014steady,chen2016generalized},  Kalman filtering \citep{chen2017maximum,8736038,liu2017maximum,wang2017maximum,ma2019unscented,liu2018maximum,song2020distributed,wang2021distributed}.  Due to the control of kernel bandwidth, the correntropy  algorithms can effectively deal with impulsive noise or outliers. 
Especially, \textcolor{black}{ the maximum correntropy criterion(MCC) has been utilized to develop robust control strategy for} Kalman filter (KF) \citep{chen2017maximum}, extended Kalman filter (EKF) \citep{8736038}, unscented Kalman filter (UKF) \citep{liu2017maximum,wang2017maximum,ma2019unscented},    cubature Kalman filter (CKF) \citep{liu2018maximum}, and distributed Kalman filter (DKF) \citep{song2020distributed,wang2021distributed} in solving non-Gaussian noises. However, the MCC algorithms mentioned above may  experience performance degradation when the estimators suffer from  more complex non-Gaussian noises, such as the multi-modal distribution noise and asymmetric noise.

Different from the MCC, the minimum error entropy (MEE) is derived by \textcolor{black}{minimizing} the Renyi’s quadratic entropy \citep{principe2010information}. According to Parzen’s estimator and Gaussian kernel, the MEE has a strong ability to model error distribution. At present, the KF \citep{8937723}, EKF \citep{8937723}, UKF \citep{dang2020robust}, and cubature information filter (CIF) \citep{li2021robust} based on the MEE criterion have been proposed for complex non-Gaussian noises, which exhibit more outstanding estimation results compared with the MCC-based filters. One  of  the   problem in MEE-based KFs  is numerical problem. As pointed out by \citep{wang2021numerically}, the matrix $\Lambda $ in \citep{8937723} is singular since the sum of its elements in each column (or row) is equal to zero, resulting in its rank \textcolor{black}{to decrease by one}. To address this issue, through a set of local solutions, a numerically stable MEE-based KF is proposed in \citep{wang2021numerically}. However, these local solutions may affect the filtering accuracy. In addition to numerical problems in the MEE-based KFs, to obtain the optimal estimation, a bias may be requiblack to eliminate the error average \citep{erdogmus2002error}.

By combining the MCC and MEE, the minimum error entropy with fiducial points (MEEF) criterion is developed in \citep{liu2006error}. Even in the case of asymmetric error distribution, putting the MCC into the MEE cost function will automatically locate the summit of error probability density function (PDF) and fix it at the origin. Recently, the CKF under MEEF criterion is proposed for INS/GPS integration \citep{9646176}, showing \textcolor{black}{better} performance than the MCC-based KFs and MEE-based KFs when the system is disturbed by complex non-Gaussian noises.  

It should be pointed out that since the performance of the KFs or its extended algorithms are deeply affected by the noise covariance matrix \citep{mohamed1999adaptive}, most of the above estimators assume that the statistics of process noise and measurement noise can be calculated accurately in advance. However, this assumption may not be established in practice, because the accurate noise statistical characteristics are arduous to be obtained for dynamic systems. Therefore, it is necessary to adopt some adaptive measures \textcolor{black}{such as Sage-Husa estimator} to update the state covariance matrix and measurement covariance matrix \citep{zhao2010adaptive,gao2015windowing}.

In this paper, inspiblack by the MEEF criterion and Sage-Husa methods, a novel  adaptive UKF is proposed to handle complex non-Gaussian noises and outliers in a system with uncertain noise distribution. \textcolor{black}{Under the MEEF criterion, the key matrix containing errors is symmetric positive definite, which is conducive to improving the numerical stability of the traditional MEE algorithm}. The contributions of this paper are listed as follows

1) {By applying the MEEF criterion to the cost function of UKF, the robust MEEF-UKF is derived.} 

2) {By means of the modified Sage-Husa method, the noise covariance matrix can be estimated in a numerically stable manner.}

3) \textcolor{black}{By maximizing the error model, the free parameters in the MEEF criteria can be optimized online.}

4) {The numerical stability and convergence of the proposed adaptive MEEF-UKF (A-MEEF-UKF) is analysed.}

The \textcolor{black}{rest} of this paper is organized as follows: Section 2 briefly reviews the UKF and MEEF. In Section 3, we derive the MEEF-UKF in detail and introduce the update method of noise covariance matrix. In Section 4, we analyze the numerical stability and convergence of the proposed algorithm. Section 5 gives the simulation results. And the conclusion of this paper is given in Section 6.

\section{Reviews}
\subsection{Unscented Kalman Filter}
The UKF is a common tool for state estimation in Gaussian nonlinear systems. Consider the following nonlinear systems
\begin{equation}
\label{e1}
\mathbf{x}_{i}=\mathbf{f}(\mathbf{x}_{i-1})+\mathbf{q}_{i-1}
\end{equation}
\begin{equation}
\label{e2}
\mathbf{y}_{i}=\mathbf{h}(\mathbf{x}_{i})+\mathbf{r}_{i}
\end{equation}
where \textcolor{black}{$\mathbf{x}_{i}$ is an n-dimensional random variable, representing the state value at time $i$.  $\mathbf{y}_{i}$ is an m-dimensional random variable, denoting the measurement at time $i$.}
$\mathbf{f}(\mathbf{x}_{i-1})$ and $\mathbf{h}(\mathbf{x}_{i})$ stand for time invariant state transition function and measurement function respectively, which is assumed to be accurately known. $\mathbf{q}_{i-1}$ denotes state noise with covariance matrix $\mathbf{Q}_{i-1}$ and $\mathbf{r}_{i}$ denotes measurement noise with covariance matrix $\mathbf{R}_{i}$. And $\mathbf{q}_{i-1}$ is not related with $\mathbf{r}_{i}$. The standard UKF contains the following two steps

1) Pblackict: Performing unscented transformation on $\hat{\mathbf{x}}_{i-1|i-1}$ and $\mathbf{P}_{i-1|i-1}$ to obtain $2n+1$ sigma points
\begin{equation}
\label{e3}
\begin{array}{l}
\chi _{i - 1|i - 1}^s = \\
\left\{ {\begin{array}{*{20}{l}}
{{{{\bf{\hat x}}}_{i - 1|i - 1}}{\rm{ }}}&{s = 0}\\
{{{{\bf{\hat x}}}_{i - 1|i - 1}} + {{(\sqrt {(n + \lambda ){\mathbf{P}_{i - 1|i - 1}}} )}_s}}&{s = 1,...,n}\\
{{\kern 1pt} {{{\bf{\hat x}}}_{i - 1|i - 1}} - {{(\sqrt {(n + \lambda ){\mathbf{P}_{i - 1|i - 1}}} )}_{s - n}}}&{s = n + 1,...,2n}
\end{array}} \right.
\end{array}
\end{equation}
where ${(\sqrt {(n + \lambda ){{\bf{P}}_{i - 1|i - 1}}} )_s}$ represents the $s$-th column of the square root of $(n + \lambda ){{\bf{P}}_{i - 1|i - 1}}$, which is realized by the square root filter \citep{van2001square}. $\lambda  = {\bar{\alpha} ^2}(n + \kappa ) - n$ is scale correction factor.

Calculating the prior state estimation $\mathbf{\hat{x}}_{i|i-1}$ and prior state covariance matrix $\mathbf{P}_{i|i-1}$  by 
\begin{equation}
\label{e4}
{{\bf{\hat x}}_{i|i - 1}} = \sum\limits_{s = 0}^{2n} {w_m^s{\bf{f}}\left( {\chi _{i - 1|i - 1}^s} \right)} 
\end{equation}
\begin{equation}
\label{e5}
{\bf{P}}_{i|i - 1} = \sum\limits_{s = 0}^{2n} {w_c^s\xi ^{s}({\xi ^{s})^T} + {{\bf{Q}}_{i - 1}}}
\end{equation}
where $\xi ^{s}=\mathbf{f}(\chi _{i-1|i-1}^{s})-\mathbf{\hat{x}}_{i|i-1}$, the weights are $w_{m}^{0}=\frac{\lambda }{n+\lambda }$, $w_{c}^{0}=\frac{\lambda }{n+\lambda }+1-\bar{\alpha} ^{2}+\bar{\beta} $, and $w_{m}^{s}=w_{c}^{s}=\frac{1 }{2(n+\lambda )}$ while $s\neq 0$.

2) Update: Performing unscented transformation on $\hat{\mathbf{x}}_{i|i-1}$ and $\mathbf{P}_{i|i-1}$ to get $2n+1$ sigma points
\begin{equation}
\label{e6}
\begin{array}{l}
\chi _{i |i - 1}^s = \\
\left\{ {\begin{array}{*{20}{l}}
{{{{\bf{\hat x}}}_{i |i - 1}}{\rm{ }}}&{s = 0}\\
{{{{\bf{\hat x}}}_{i |i - 1}} + {{(\sqrt {(n + \lambda ){\mathbf{P}_{i |i - 1}}} )}_s}}&{s = 1,...,n}\\
{{\kern 1pt} {{{\bf{\hat x}}}_{i |i - 1}} - {{(\sqrt {(n + \lambda ){\mathbf{P}_{i |i - 1}}} )}_{s - n}}}&{s = n + 1,...,2n}
\end{array}} \right.
\end{array}
\end{equation}
The square root of $(n + \lambda ){{\bf{P}}_{i|i - 1}}$ is also realized by the square root filter \citep{van2001square}.

Calculating the prior measurement $\mathbf{\hat{y}}_{i|i-1}$, cross-covariance matrix $\mathbf{P}_{xy,i|i-1}$, and measurement covariance matrix $\mathbf{P}_{yy,i|i-1}$  by 
\begin{equation}
\label{e7}
{{\bf{\hat y}}_{i|i - 1}} = \sum\limits_{s = 0}^{2n} {w_m^s{\bf{f}}\left( {\chi _{i |i - 1}^s} \right)} 
\end{equation}
\begin{equation}
\label{e8}
{\bf{P}}_{xy,i|i - 1} = \sum\limits_{s = 0}^{2n} {w_c^s\xi ^{s}({\zeta  ^{s})^T}}
\end{equation}
\begin{equation}
\label{e9}
{\bf{P}}_{yy,i|i - 1} = \sum\limits_{s = 0}^{2n} {w_c^s\zeta  ^{s}({\zeta  ^{s})^T}}+\mathbf{R}_{i}
\end{equation}
where $\zeta  ^{s}=\mathbf{h}(\chi _{i|i-1}^{s})-\mathbf{\hat{y}}_{i|i-1}$.

The posteriori state estimation $\mathbf{\hat{x}}_{i|i}$ and state covariance $\mathbf{P}_{i|i}$ are updated by 
\begin{equation}
\label{e10}
\mathbf{\hat{x}}_{i|i}=\mathbf{\hat{x}}_{i|i-1}+\mathbf{K}_{i}(\mathbf{y}_{i}-\mathbf{\hat{y}}_{i|i-1})
\end{equation}
\begin{equation}
\label{e11}
\mathbf{P}_{i|i}=\mathbf{P}_{i|i-1}-\mathbf{K}_{i}\mathbf{P}_{yy,i|i-1}\mathbf{K}_{i}^{T}
\end{equation}
where the gain matrix is $\mathbf{K}_{i}=\mathbf{P}_{xy,i|i-1}\mathbf{P}_{yy,i|i-1}^{-1}$.

\subsection{Minimum Error Entropy with Fiducial Points}
Given two random variables $X$ and $Y$, the correntropy is defined by 
\begin{equation}
\label{e12}
V_{1}(e)=E[\kappa _{\sigma }(e)]=\int \kappa _{\sigma }(e)p(e)de
\end{equation}
where the error information $e=X-Y$ has the PDF $p(e)$. And $\kappa _{\sigma }(e)$ represents kernel function, in which the Gaussian kernel $ G_{\sigma }(e)=exp(-e^{2}/(2\sigma ^{2})) $ is used in this paper.   $\sigma > 0 $ presents kernel width. Since only \textcolor{black}{the} limited samples $\left \{ e_{k} \right \}_{k=1}^{N}$ are known, \eqref{e12} can be estimated as
\begin{equation}
\label{e13}
\hat{V}_{1}(e)=\frac{1}{N}\sum_{k=1}^{N}G_{\sigma }(e_{k})
\end{equation}
The MEE is proposed by \textcolor{black}{minimizing} the notable quadratic Renyi’s entropy $H_{2}(e)=-logV_{2}(e)$ \citep{liu2007correntropy,principe2010information,erdogmus2002error}, where the quadratic error information potential $V_{2}(e)$ is defined by
\begin{equation}
\label{e14}
V_{2}(e)=E[p(e)]=\int p(e)p(e)de
\end{equation}
According to Parzen's estimator, the PDF $p(e)$ can be calculated by $\hat{p}(e)=\frac{1}{N}\sum_{k=1}^{N}G_{\sigma }(e-e_{k})$. 
Then \eqref{e14} can be estimated by 
\begin{equation}
\label{e15}
\hat{V}_{2}(e)=\frac{1}{N^{2}}\sum_{j=1}^{N}\sum_{k=1}^{N}G_{\sigma }(e_{j}-e_{k})
\end{equation}
\textcolor{black}{From the surface of the loss function \citep{liu2006error, 9646176},  the function can obtain the extreme value when the errors are on the line of $\frac{\pi}{4}$ and $\frac{5\pi}{4}$, resulting in} the errors may not be located at the origin after the problem is optimized. In order to set the errors at zero in an automatic way, the MEEF criterion takes into account adding the MCC to MEE, which is given by \citep{liu2006error}
\begin{equation}
\label{e16}
J=\tau  \sum_{k=1}^{N}G_{\sigma _{ 1 }}(e_{k})+(1-\tau  )\sum_{j=1}^{N}\sum_{k=1}^{N}G_{\sigma_{ 2 }}(e_{j}-e_{k})
\end{equation}
where \textcolor{black}{ $J$ is the cost function when the model adopts the MEEF as the optimization criterion}. The fusion factor $0\leq \tau \leq 1$ determines how many fiducial points are placed at the origin. It is obvious that when $\tau $ is equal to zero, the MEEF will degenerate to MEE, and when $\tau $ is equal to one, the MEEF will degenerate to MCC.

\section{Proposed Algorithm}

\subsection{Unscented Kalman Filter under MEEF}
To process state error and measurement error simultaneously, it is necessary to construct a  linear batch regression model. Traditionally, the linear  model of \eqref{e2} in EKF is realized by Taylor's first-order approximation  \citep{8736038,8937723}, and in UKF is achieved by statistical linear technique \citep{liu2017maximum,ma2019unscented,dang2020robust}.
\begin{equation}
\label{e17}
\begin{bmatrix}
\mathbf{\hat{x}}_{i|i-1}
\\ \mathbf{y}_{i}-\mathbf{\hat{y}}_{i|i-1}+\mathbf{S}_{i}\mathbf{\hat{x}}_{i|i-1}
\end{bmatrix}=
\begin{bmatrix}
\mathbf{I}_{n}
\\ \mathbf{S}_{i}
\end{bmatrix}
\mathbf{x}_{i}+
\begin{bmatrix}
\delta _{i}\\ 
\mathbf{r}_{i}
\end{bmatrix}
\end{equation}
where $\mathbf{S}_{i}=(\mathbf{P}_{i|i-1}^{-1}\mathbf{P}_{xy,i|i-1})^{T}$ denotes slope matrix, $\mathbf{I}_{n}\in\mathbb{R}^{n\times n}$ is an identity matrix, and $\delta _{i}=\mathbf{\hat{x}}_{i|i-1}-\mathbf{x}_{i}$ is the state pblackiction error. Define $\varepsilon _{i}= [\delta _{i}^{T},\mathbf{r}_{i}^T]^{T}$, the covariance matrix of $\varepsilon _{i}$ can be calculated by
\begin{equation}
\label{e18}
E[\varepsilon _{i}\varepsilon _{i}^{T}]=\mathbf{\Xi }_{i}\mathbf{\Xi }_{i}^{T}=\begin{bmatrix}
\mathbf{\Xi }_{p,i|i-1}\mathbf{\Xi }_{p,i|i-1}^{T} & \mathbf{0}\\ 
\mathbf{0} & \mathbf{\Xi }_{r,i}\mathbf{\Xi }_{r,i}^{T}
\end{bmatrix}
\end{equation}
where $\mathbf{\Xi }_{i}$, $\mathbf{\Xi }_{p,i|i-1}$, and $\mathbf{\Xi }_{r,i}$ are the Cholesky factorization  of $E[\varepsilon _{i}\varepsilon _{i}^{T}]$, $\mathbf{P}_{i|i-1}$, and $\mathbf{R}_{i}$, respectively. \textcolor{black}{Here, the existence of the inverse operation and Cholesky factorization  of $\mathbf{P}_{i|i-1}$ are the same as that of UKF.}

Multiplying $\mathbf{\Xi }_{i}^{-1}$ to the left of \eqref{e17}
\begin{equation}
\label{e19}
{{\bf{z}}_i} = {{\bf{A}}_i}{{\bf{x}}_i} + {{\bf{e}}_i}
\end{equation}
where 
\begin{equation}
\label{e20}
\left\{ {\begin{array}{*{20}{l}}
{{{\bf{z}}_i} = {\bf{\Xi }}_i^{ - 1}{{\left[ {{\bf{\hat x}}_{i|i - 1}^T,{{({{\bf{y}}_i} - {{{\bf{\hat y}}}_{i|i - 1}} + {{\bf{S}}_i}{{{\bf{\hat x}}}_{i|i - 1}})}^T}} \right]}^T}}\\
{{{\bf{A}}_i} = {\bf{\Xi }}_i^{ - 1}{{[{{\bf{I}}_n},{\bf{S}}_i^T]}^T}}\\
{{{\bf{e}}_i} = {\bf{\Xi }}_i^{ - 1}{{\bf{\varepsilon }}_i}}
\end{array}} \right.
\end{equation}
with $\mathbf{z}_{i}=[z_{i,1},z_{i,2},\cdots,z_{i,N}]^{T}$, $\mathbf{A}_{i}=[\mathbf{a}_{i,1}^{T},\mathbf{a}_{i,2}^{T},\cdots,\mathbf{a}_{i,N}^{T}]^{T}$, $\mathbf{e}_{i}=[e_{i,1},e_{i,2},\cdots ,e_{i,N}]^{T}$, and $N=n+m$. Note that $E[\mathbf{e}_{i}\mathbf{e}_{i}^{T}]=\mathbf{I}_{N}$, the residual error $\mathbf{e}_{i}$ is white.

\textcolor{black}{In the convenience of converting the MEEF-UKF  to the MEE-UKF or MCC-UKF by taking different values of $\tau$}, the cost function of MEEF-UKF can be defined by
\begin{equation}
\label{e21}
\begin{array}{l}
J({{\bf{x}}_i})\\
{\rm{ = }}\tau (2\sigma _1^2)\sum\limits_{k = 1}^N {{G_{{\sigma _1}}}\left( {{e_{i,k}}} \right)}  + (1 - \tau )\sigma _2^2\sum\limits_{j = 1}^N {\sum\limits_{k = 1}^N {{G_{{\sigma _2}}}\left( {{e_{i,j}} - {e_{i,k}}} \right)} } 
\end{array}
\end{equation}
The optimal solution $\mathbf{\hat{x}}_{i|i}$  is obtained by maximizing \eqref{e21}, i.e. $\mathbf{\hat{x}}_{i|i}=\mathop {\textup{arg\ max} }\limits_{{{\bf{x}}_i}} \left\{ {J({{\bf{x}}_i})} \right\}$. Let the derivative of $J(\mathbf{x}_{i})$  with respect to  $\mathbf{x}_{i}$ be zero
\begin{equation}
\label{e22}
\begin{aligned}
\frac{\partial }{{{{\bf{x}}_i}}}J\left( {{{\bf{x}}_i}} \right)  &= 2\tau \sum\limits_{k = 1}^N {{G_{{\sigma _1}}}\left( {{e_{i,k}}} \right)} {\bf{a}}_{i,k}^T{e_{i,k}}\\
 &+ 2(1 - \tau )\sum\limits_{j = 1}^N {\sum\limits_{k = 1}^N {{G_{{\sigma _2}}}\left( {{e_{i,j}} - {e_{i,k}}} \right)} } {\bf{a}}_{i,j}^T{e_{i,k}}\\
 &- 2(1 - \tau )\sum\limits_{j = 1}^N {\sum\limits_{k = 1}^N {{G_{{\sigma _2}}}\left( {{e_{i,j}} - {e_{i,k}}} \right)} } {\bf{a}}_{i,k}^T{e_{i,k}}\\
 &=0
\end{aligned}
\end{equation}
Simplifying \eqref{e22} as
\begin{equation}
\label{e23}
\tau {\bf{A}}_i^T{{\bf{\Lambda  }}_i}{{\bf{e}}_i} + (1 - \tau ){\bf{A}}_i^T({{\bf{\Psi }}_i} - {{\bf{\Omega }}_i}){{\bf{e}}_i} = 0
\end{equation}
where
\begin {small}
\begin{equation}
\label{e24}
\left\{ {\begin{array}{*{20}{l}}
{{{\bf{\Lambda }}_i} = {\rm{diag}}\left( {{G_{{\sigma _1}}}\left( {{e_{i,1}}} \right),{G_{{\sigma _1}}}\left( {{e_{i,2}}} \right), \cdots ,{G_{{\sigma _1}}}\left( {{e_{i,N}}} \right)} \right)}\\
{{{\bf{\Psi }}_i} = {\rm{diag}}\left( {\sum\limits_{j = 1}^N {{G_{{\sigma _2}}}\left( {{e_{i,1}} - {e_{i,j}}} \right)} , \cdots ,\sum\limits_{j = 1}^N {{G_{{\sigma _2}}}\left( {{e_{i,N}} - {e_{i,j}}} \right)} } \right)}\\
{{{\left\{ {{{\bf{\Omega }}_i}} \right\}}_{jk}} = {G_{{\sigma _2}}}\left( {{e_{i,j}} - {e_{i,k}}} \right)}
\end{array}} \right.
\end{equation}
\end{small}
where ${{\left\{ {{{\bf{\Omega }}_i}} \right\}}_{jk}}$ denotes the element in row $j$ and column $k$ of ${{{\bf{\Omega }}_i}}$. Through \eqref{e19},  \eqref{e23} can be blackuced to
\begin{equation}
\label{e25}
{{\bf{x}}_i} = {\left( {{\bf{A}}_i^T\mathbf{\Phi} _{i}{{\bf{A}}_i}} \right)^{ - 1}}{\bf{A}}_i^T\mathbf{\Phi} _{i}{{\bf{z}}_i}
\end{equation}
where 
\begin{equation}
\label{e26}
\mathbf{\Phi} _{i}=\tau \mathbf{\Lambda} _{i}+(1-\tau)(\mathbf{\Psi}_{i}-\mathbf{\Omega}_{i}  )
\end{equation}
It can be seen that the MEEF-UKF will degenerate into standard UKF when $\mathbf{\Phi} _{i}=\mathbf{I}_{N}$ ($\sigma _{1}\to \infty $ and $\tau=1$), will turn into the MCC-UKF when $\mathbf{\Phi} _{i}=\mathbf{\Lambda} _{i}$ ($\tau=1$), and will become to the MEE-UKF when $\mathbf{\Phi} _{i}=\mathbf{\Psi}_{i}-\mathbf{\Omega}_{i}$ ($\tau=0$). 

Blocking $\mathbf{\Phi} _{i}$ as 
\begin{equation}
\label{e27}
{{\bf{\Phi }}_i} = \left[ {\begin{array}{*{20}{c}}
{{{\bf{\Phi }}_{xx,i}}}&{{{\bf{\Phi }}_{yx,i}}}\\
{{{\bf{\Phi }}_{xy,i}}}&{{{\bf{\Phi }}_{yy,i}}}
\end{array}} \right]
\end{equation}
where $\mathbf{\Phi} _{xx,i}\in \mathbb{R}^{n\times n}$, $\mathbf{\Phi} _{xy,i}\in \mathbb{R}^{m\times n}$, $\mathbf{\Phi} _{yx,i}\in \mathbb{R}^{n\times m}$, and $\mathbf{\Phi} _{yy,i}\in \mathbb{R}^{m\times m}$.

Since $\mathbf{\Phi} _{i}$ is related to $\mathbf{x}_i$,  \eqref{e25} can be expressed as $\mathbf{x}_{i}=\mathbf{g}(\mathbf{x}_{i})$, which can be solved by fixed-point iteration, i.e. $\mathbf{\hat{x}}_{i|i}^{t+1}=\mathbf{g}(\mathbf{\hat{x}}_{i|i}^t)$.  According to \eqref{e20},  \eqref{e27}, and the matrix inverse lemma \citep{8937723,dang2020robust}, the \eqref{e25} can be rewritten as
\begin{equation}
\label{e28}
{{\bf{\hat x}}_{i|i}} = {{\bf{\hat x}}_{i|i - 1}} + {{\bf{\bar K}}_i}\left( {{{\bf{y}}_i} - {{{\bf{\hat y}}}_{i|i - 1}}} \right)
\end{equation}
where the gain matrix ${{\bf{\bar K}}_i}$ is
\begin{equation}
\label{e29}
\begin{aligned}
{{{\bf{\bar K}}}_i} & = {\left( {{{{\bf{\bar P}}}_{xx,i}} + {\bf{S}}_i^T{{{\bf{\bar P}}}_{xy,i}} + {{{\bf{\bar P}}}_{yx,i}}{{\bf{S}}_i} + {\bf{S}}_i^T{{{\bf{\bar R}}}_{yy,i}}{{\bf{S}}_i}} \right)^{ - 1}}\\
& \times \left( {{{{\bf{\bar P}}}_{yx,i}} + {\bf{S}}_i^T{{{\bf{\bar R}}}_{yy,i}}} \right){\kern 1pt} 
\end{aligned}
\end{equation}
with
\begin{equation}
\label{e30}
\left\{ {\begin{array}{*{20}{l}}
{{{{\bf{\bar P}}}_{xx,i}} = {{\left( {{\bf{\Xi }}_{p,i|i - 1}^{ - 1}} \right)}^T}{{\bf{\Phi }}_{xx,i}}{\bf{\Xi }}_{p,i|i - 1}^{ - 1}}\\
{{{{\bf{\bar P}}}_{xy,i}} = {{\left( {{\bf{\Xi }}_{r,i}^{ - 1}} \right)}^T}{{\bf{\Phi }}_{xy,i}}{\bf{\Xi }}_{p,i|i - 1}^{ - 1}}\\
{{{{\bf{\bar P}}}_{yx,i}} = {{\left( {{\bf{\Xi }}_{p,i|i - 1}^{ - 1}} \right)}^T}{{\bf{\Phi }}_{yx,i}}{\bf{\Xi }}_{r,i}^{ - 1}}\\
{{{{\bf{\bar R}}}_{yy,i}} = {{\left( {{\bf{\Xi }}_{r,i}^{ - 1}} \right)}^T}{{\bf{\Phi }}_{yy,i}}{\bf{\Xi }}_{r,i}^{ - 1}}
\end{array}} \right.
\end{equation}
It should be noted that the inverse operation in \eqref{e29} will be replaced by pseudo-inverse operation to enhance numerical stability.
Defining $\mathbf{P}_{p,i}  ={{{{\bf{\bar P}}}_{xx,i}} + {\bf{S}}_i^T{{{\bf{\bar P}}}_{xy,i}} + {{{\bf{\bar P}}}_{yx,i}}{{\bf{S}}_i} + {\bf{S}}_i^T{{{\bf{\bar R}}}_{yy,i}}{{\bf{S}}_i}}$, the pseudo-inverse of $\mathbf{P}_{p,i}$ is expressed as $\mathbf{P}_{p,i}^+$.  One way to calculate $\mathbf{P}_{p,i}^+$ is to employ SVD, which is realized by the following steps: Performing SVD on $\mathbf{P}_{p,i}$, we have
\begin{equation} \label{ee1}
\mathbf{P}_{p,i}=\mathbf{U}\mathbf{D}\mathbf{V}^{T}
\end{equation}
where $\mathbf{U}\in \mathbb{R}^{n\times n}$ and $\mathbf{V}\in \mathbb{R}^{n\times n}$ are orthogonal matrix, $\mathbf{D}\in \mathbb{R}^{n\times n}$ is a diagonal matrix composed of singular values. Invert all non-zero elements in $\mathbf{D}$ to obtain $\mathbf{D}^+$, and then the  $\mathbf{P}_{p,i}^+$ can be calculated by
\begin{equation} \label{ee2}
\mathbf{P}_{p,i}^+=\mathbf{V}\mathbf{D}^+\mathbf{U}^{T}
\end{equation}
It can be seen that $\mathbf{P}_{p,i}^+=\mathbf{P}_{p,i}^-$ when the matrix $\mathbf{P}_{p,i}$ is non-singular. Even if $\mathbf{P}_{p,i}$ is singular, the pseudo-inverse $\mathbf{P}_{p,i}^+$ still exists. The good numerical stability of $\mathbf{P}_{p,i}^+$ has been reflected in \citep{kulikov2019numerical}. Then \eqref{e29} can be expressed as 
\begin{equation} \label{ee3}
{{\bf{\bar K}}_i} = \mathbf{P}_{p,i}^+  \times \left( {{{{\bf{\bar P}}}_{yx,i}} + {\bf{S}}_i^T{{{\bf{\bar R}}}_{yy,i}}} \right)
\end{equation}

The state covariance matrix $\mathbf{P}_{i|i}=E[(\mathbf{x}_{i}-\mathbf{\hat{x}}_{i|i})(\mathbf{x}_{i}-\mathbf{\hat{x}}_{i|i})^{T}]$ is given by 
\begin{equation}
\label{e31}
{{\bf{P}}_{i|i}} = \left( {{{\bf{I}}_n} - {{{\bf{\bar K}}}_i}{{\bf{S}}_i}} \right){{\bf{P}}_{i|i - 1}}{\left( {{{\bf{I}}_n} - {{{\bf{\bar K}}}_i}{{\bf{S}}_i}} \right)^T} + {{\bf{\bar K}}_i}{{\bf{R}}_i}{\bf{\bar K}}_i^T
\end{equation}

\textit{Remark 1}: According to the performance surface of MEEF in \citep{liu2006error,9646176}, the peak of MEEF is fixed at the origin due to the introduction of MCC. In fact, the MEEF-UKF inherits the advantages of the MEE and MCC. On the one hand, MEE has a strong ability to model the error distribution through the Parzen's estimator and Gaussian kernel in \eqref{e15}. On the other hand, through Gaussian kernels $G_{\sigma _{1}}(e_{i,k})$ and $G_{\sigma _{2}}(e_{i,j}-e_{i,k})$, both of MCC and MEE can cope with abnormal $\mathbf{e}_{i}$ caused by large outliers or noise \citep{8793125,7795189}. In addition, the matrix $\mathbf{\Phi} _{i}$ in MEEF-UKF is symmetric positive definite (see Section 4 for detailed proof), which has superior stability to the MEE-UKF.

\subsection{Estimation of Noise Covariance Matrix}
In a practical system, the process noise covariance matrix $\mathbf{Q}_i$ and measurement noise covariance matrix $\mathbf{R}_i$ are deeply affected by the dynamic change of the system, which makes it difficult for the algorithm to estimate under a certain noise condition. In order to further augment the estimation accuracy of MEEF-UKF, $\mathbf{Q}_i$ and $\mathbf{R}_i$ need to be dynamically estimated at each time. For a nonlinear dynamic system, \textcolor{black}{according to \citep{zhao2010adaptive,gao2015windowing}, $\mathbf{Q}_i$ and $\mathbf{R}_i$ can be evaluated by}
\textcolor{black}{
\begin{equation}
\label{eee32}
\begin{aligned}
\mathbf{\hat{Q}}_{i} & = (1-d_{i})\mathbf{\hat{Q}}_{i-1} \\
& +d_{i}\left (\mathbf{{K}}_{i}\mathbf{\tilde{y}}_{i}\mathbf{\tilde{y}}_{i}^{T}\mathbf{{K}}_{i}^{T}+ \mathbf{P}_{i|i}-\sum\limits_{s = 0}^{2n} {w_c^s(\mathbf{f}(\chi _{i-1|i-1}^{s})-\mathbf{\hat{x}}_{i|i-1})(\mathbf{f}(\chi _{i-1|i-1}^{s})-\mathbf{\hat{x}}_{i|i-1})^T}  \right )
\end{aligned}
\end{equation}
\begin{equation}
\label{eee33}
\mathbf{\hat{R}}_{i+1}=(1-d_{i})\mathbf{\hat{R}}_{i}+d_{i}\left (\mathbf{\tilde{y}}_{i}\mathbf{\tilde{y}}_{i}^{T}+  \sum\limits_{s = 0}^{2n} {w_c^s(\mathbf{h}(\chi _{i|i-1}^{s})-\mathbf{\hat{y}}_{i|i-1})({\mathbf{h}(\chi _{i|i-1}^{s})-\mathbf{\hat{y}}_{i|i-1})^T}}  \right )
\end{equation}
where $d_{i}={(1-b)}/{(1-b^{i+1})}$, the constant $b$ is the forgetting factor, determining the proportion of data in the past.} And the $ \mathbf{\tilde{y}}_{i}={{\bf{y}}_i} - {{{\bf{\hat y}}}_{i|i - 1}} $ represents innovation vector. \textcolor{black}{In the MEEF-UKF, the gain matrix is ${{\bf{\bar K}}_i}$, which can take over the ${\bf{ K}}_i$ in \eqref{eee32}. Combining \eqref{e5}, \eqref{e9}, \eqref{eee32}, and \eqref{eee33}, we have}
\begin{equation}
\label{e32}
\begin{aligned}
\mathbf{\hat{Q}}_{i} & = (1-d_{i})\mathbf{\hat{Q}}_{i-1} \\
& +d_{i}\left (\mathbf{\bar{K}}_{i}\mathbf{\tilde{y}}_{i}\mathbf{\tilde{y}}_{i}^{T}\mathbf{\bar{K}}_{i}^{T}+ \mathbf{P}_{i|i}-\mathbf{P}_{i|i-1}+\mathbf{\hat{Q}}_{i-1} \right )
\end{aligned}
\end{equation}
\begin{equation}
\label{e33}
\mathbf{\hat{R}}_{i+1}=(1-d_{i})\mathbf{\hat{R}}_{i}+d_{i}\left (\mathbf{\tilde{y}}_{i}\mathbf{\tilde{y}}_{i}^{T}- \mathbf{P}_{yy,i|i-1}+\mathbf{\hat{R}}_{i}  \right )
\end{equation}

From \eqref{e32}, when the state vector $\mathbf{x}_i$ \textcolor{black}{fluctuates greatly}, the positive definiteness of the estimated process noise covariance matrix $\mathbf{\hat{Q}}_i$ is difficult to be guaranteed, which may lead to the failure of the estimator to perform Cholesky factorization . Considering that $\mathbf{Q}_i$ and $\mathbf{R}_i$ are diagonal matrices, since $\mathbf{q}_i$ and $\mathbf{r}_i$ are uncorrelated noises, their estimation can be revised as
\begin{equation}
\label{e34}
\mathbf{\hat{Q}}_{i}=\sqrt{diag(\mathbf{\hat{Q}}_{i}\mathbf{\hat{Q}}_{i}^{T})}
\end{equation}
\begin{equation}
\label{e35}
\mathbf{\hat{R}}_{i+1}=\sqrt{diag(\mathbf{\hat{R}}_{i+1}\mathbf{\hat{R}}_{i+1}^{T})}
\end{equation}
where the operation $diag(\cdot )$  means to extract the diagonal elements of the original matrix as a new diagonal matrix. Obviously, the estimates of $\mathbf{\hat{Q}}_i$ and $\mathbf{\hat{R}}_{i+1}$ are positive definite.

The A-MEEF-UKF is summarized in Algorithm 1.
\begin{algorithm}
\caption{A-MEEF-UKF}\label{am1}
\begin{algorithmic}
\STATE 
\STATE \textbf{Initiation }: Set initial $\mathbf{\hat{x}}_{0|0}$, $\mathbf{P}_{0|0}$, $\mathbf{\hat{Q}}_0$, $\mathbf{\hat{R}}_{1}$, set kernel width $\sigma _{1}$, $\sigma _{2}$, and threshold $\gamma $. 
\STATE \textbf{For} $i=1,2,3...$ 
\STATE \hspace{0.5cm} Calculate $\mathbf{\hat{x}}_{i|i-1}$, $\mathbf{P}_{i|i-1}$, $\mathbf{\hat{y}}_{i|i-1}$, $\mathbf{P}_{xy,i|i-1}$,  and $\mathbf{P}_{yy,i|i-1}$ through \eqref{e3}-\eqref{e9}. Obtain slope matrix $\mathbf{S}_{i}=(\mathbf{P}_{i|i-1}^{-1}\mathbf{P}_{xy,i|i-1})^{T}$. Perform Cholesky factorization  on $\mathbf{P}_{i|i-1}$ and $\mathbf{\hat{R}}_{i}$ to get $\mathbf{\Xi} _{p,i|i-1}$, $\mathbf{\Xi} _{r,i}$ and $\mathbf{\Xi} _{i}$. Calculate  $\mathbf{z}_{i}$ and $\mathbf{A}_{i} $ through \eqref{e20}
\STATE \hspace{0.5cm} To solve \eqref{e25}, the fixed-point iteration is given by the following steps: 
\STATE \hspace{0.5cm} Set initial iterative values as $\mathbf{\hat{x}}_{i|i}^{0}=\mathbf{\hat{x}}_{i|i-1}$ and $t=1$; Then, calculate residual error $\mathbf{e}_{i}^{t}$ by
\begin{equation}
\label{e36}
\mathbf{e}_{i}^{t}=\mathbf{z}_{i}-\mathbf{A}_{i}\mathbf{\hat{x}}_{i|i}^{t-1}
\end{equation}
\STATE \hspace{0.5cm} Obtain ${{{\bf{\Phi }}_{xx,i}}}$, ${{{\bf{\Phi }}_{xy,i}}}$, ${{{\bf{\Phi }}_{yx,i}}}$, and ${{{\bf{\Phi }}_{yy,i}}}$ through \eqref{e24}, \eqref{e26}, and \eqref{e27}. Calculate $\mathbf{\bar{K}}_{i}$ through \eqref{e29} - \eqref{ee3}. Calculate $\mathbf{\hat{x}}_{i|i}^{t}$ by
\begin{equation}
\label{e37}
{{\bf{\hat x}}_{i|i}^t} = {{\bf{\hat x}}_{i|i - 1}} + {{\bf{\bar K}}_i}\left( {{{\bf{y}}_i} - {{{\bf{\hat y}}}_{i|i - 1}}} \right)
\end{equation}
\STATE \hspace{0.5cm} \textbf{If} 
$||\mathbf{\hat{x}}_{i|i}^{t}-\mathbf{\hat{x}}_{i|i}^{t-1}||/||\mathbf{\hat{x}}_{i|i}^{t-1}||\leq  \gamma $ holds, set $\mathbf{\hat{x}}_{i|i}=\mathbf{\hat{x}}_{i|i}^{t}$, and go to the next step. \textbf{Otherwise}, set $t=t+1$, and go back to \eqref{e36}.
\STATE \hspace{0.5cm} Update $\mathbf{P}_{i|i}$, $\mathbf{\hat{Q}}_{i}$, and $\mathbf{\hat{R}}_{i+1}$ by \eqref{e31}, \eqref{e32}, and \eqref{e33}. Revise $\mathbf{\hat{Q}}_{i}$ and $\mathbf{\hat{R}}_{i+1}$ by \eqref{e34} and \eqref{e35}.
\STATE \textbf{End For}
\end{algorithmic}
\label{AM1}
\end{algorithm}

\subsection{ \textcolor{black}{Determination of free parameters in A-MEEF-UKF}}
\textcolor{black}{The free parameters in MEEF criterion include fusion factor $\tau$, kernel width $\sigma _{1}$ and kernel width $\sigma _{2}$. How to select these parameters is very important, which determines the estimation accuracy and robustness of A-MEEF-UKF. First, from the perspective of accuracy, these free parameters can be obtained by maximizing the MEEF model in \eqref{e21}.
\begin{equation}
\label{ee38}
(\tau ^{\circ },\sigma _{1}^{\circ },\sigma _{2}^{\circ })  = \mathop{\textup{arg\, max}}\limits_{\tau \in S_{\tau},\sigma _{1} \in S_{\sigma _{1}},\sigma _{2} \in S_{\sigma _{2}} } V_{MEEF} 
\end{equation}
where $S_{\tau}$, $S_{\sigma _{1}}$, and $S_{\sigma _{2}}$ denote the value set of $\tau$, $\sigma_1$, and $\sigma_2$, respectively. The $V_{MEEF}$ is 
\begin{equation}
\label{ee39}
\begin{aligned}
  {{V}_{MEEF}} &=\tau (2\sigma _{1}^{2})E\left[ {{G}_{{{\sigma }_{1}}}}(e) \right]+(1-\tau )\sigma _{2}^{2}E\left[ p(e) \right] \\ 
 & =\int_{-\infty }^{\infty }{\left[ \tau (2\sigma _{1}^{2}){{G}_{{{\sigma }_{1}}}}(x)+(1-\tau )\sigma _{2}^{2}{{p}_{e}}(x) \right]{{p}_{e}}(x)dx} \\ 
 & =\frac{1}{2}\int_{-\infty }^{\infty }{{{\left[ \tau (2\sigma _{1}^{2}){{G}_{{{\sigma }_{1}}}}(x)+(1-\tau )\sigma _{2}^{2}{{p}_{e}}(x) \right]}^{2}}dx}+ \\
 & \frac{1}{2}\int_{-\infty }^{\infty }{{{\left[ {{p}_{e}}(x) \right]}^{2}}dx}-\frac{1}{2}\int_{-\infty }^{\infty }{{{\left[ \tau (2\sigma _{1}^{2}){{G}_{{{\sigma }_{1}}}}(x)+(1-\tau )\sigma _{2}^{2}{{p}_{e}}(x)-{{p}_{e}}(x) \right]}^{2}}dx} \\ 
\end{aligned}
\end{equation}
When the PDF of the error is fixed at discrete time $i$, the maximum value of \eqref{ee39} is only determined by the third term \citep{chen2021multikernel,9646176}. Then free parameters can be obtained by minimizing the following formula
\begin{equation}
\begin{aligned}
\label{ee40}
(\tau ^{\circ },\sigma _{1}^{\circ },\sigma _{2}^{\circ })  &= \mathop{\textup{arg\, min}}\limits_{\tau \in S_{\tau},\sigma _{1} \in S_{\sigma _{1}},\sigma _{2} \in S_{\sigma _{2}} } \int_{-\infty }^{\infty }{{{\left[ \tau (2\sigma _{1}^{2}){{G}_{{{\sigma }_{1}}}}(x)+(1-\tau )\sigma _{2}^{2}{{p}_{e}}(x)-{{p}_{e}}(x) \right]}^{2}}dx}\\
&=  \mathop{\textup{arg\, min}}\limits_{\tau \in S_{\tau},\sigma _{1} \in S_{\sigma _{1}},\sigma _{2} \in S_{\sigma _{2}} }  \int_{-\infty }^{\infty }{{{\left[ \tau (2\sigma _{1}^{2}){{G}_{{{\sigma }_{1}}}}(x)+(1-\tau )\sigma _{2}^{2}{{p}_{e}}(x) \right]}^{2}}dx}\\ &- 2\int_{-\infty }^{\infty }{{{\left[ \tau (2\sigma _{1}^{2}){{G}_{{{\sigma }_{1}}}}(x)+(1-\tau )\sigma _{2}^{2}{{p}_{e}}(x) \right]}}p_e(x)dx}  \\
\end{aligned}
\end{equation}
Since $p_e(x)$ can be estimated as $\hat{p}_e(x)=\frac{1}{N}\sum_{k=1}^{N}G_{\sigma }(x-e_{k})$, \eqref{ee40} can be expressed as
\begin{equation}
\begin{aligned}
\label{ee41}
(\tau ^{\circ },\sigma _{1}^{\circ },\sigma _{2}^{\circ })  &= \mathop{\textup{arg\, min}}\limits_{\tau \in S_{\tau},\sigma _{1} \in S_{\sigma _{1}},\sigma _{2} \in S_{\sigma _{2}} }  \overline{\boldsymbol{\tau }}^{T}\overline{\boldsymbol{G }}\overline{\boldsymbol{\tau }}-2\overline{\boldsymbol{\tau }}^{T}\overline{\boldsymbol{h }}
\end{aligned}
\end{equation}
where $ \overline{\boldsymbol{\tau }}=[2\tau\sigma _{1}^{2},(1-\tau)\sigma _{2}^{2}/N,...,(1-\tau)\sigma _{2}^{2}/N]^T\in \mathbb{R}^{(N+1)\times 1}$,  $\overline{\boldsymbol{h }}=\frac{1}{N}\sum_{k=1}^{N}\boldsymbol{G }_{\sigma _{1},\sigma _{2}}(e_{k})$ with $\boldsymbol{G }_{\sigma _{1},\sigma _{2}}(e_{k})=[G_{\sigma _{1}}(e_{k}),G_{\sigma _{2}}(e_{k}-e_{1}),...,G_{\sigma _{2}}(e_{k}-e_{N})]$, and $\overline{\boldsymbol{G }} = \int_{-\infty }^{\infty}\boldsymbol{G }^{T}_{\sigma _{1},\sigma _{2}}(x) \times  \boldsymbol{G }_{\sigma _{1},\sigma _{2}}(x)dx$. 
The  fusion  vector $\overline{\tau}   $  is  optimized by maximizing $\overline{\boldsymbol{\tau }}^{T}\overline{\boldsymbol{G }}\overline{\boldsymbol{\tau }}-2\overline{\boldsymbol{\tau }}^{T}\overline{\boldsymbol{h }}$:
\begin{equation}
\begin{aligned}
\label{ee42}
\overline{\boldsymbol{\tau }} ^\circ   =(\overline{\boldsymbol{G }}+\iota \mathbf{I})^{-1}\overline{\boldsymbol{h }}
\end{aligned}
\end{equation}
where $\iota$ is a regularization  parameter to avoid numerical problem. Then we can cross iterate $L$ times from a given finite set to search for the optimal parameters. The process of determining free parameters is summarized as Algorithm 2.}

\textcolor{black}{\textit{Remark 2}: Because Algorithm 2 uses alternating iteration to obtain the optimal free parameters, which brings a lot of computational burden. By formulating the finite set $S_{\sigma_1}$ and $S_{\sigma_2}$, the computational burden can be well avoided. On the other hand, setting $S_{\sigma_1}$ and $S_{\sigma_2}$ can improve the performance of the algorithm, because too large kernel width will blackuce the robustness of the algorithm against outliers \citep{8793125,7795189}, and too small kernel width will lead to filter  divergence(See the next Section).}

\begin{algorithm}
\caption{\textcolor{black}{Determination of the free parameters}}\label{am2}
\begin{algorithmic}
\STATE 
\STATE \textbf{Initiation }: Input error sample $\mathbf{e}_{i}^{t}$ in \eqref{e36}.  Set the  proper finite set $S_{\sigma _{1}}$, $S_{\sigma _{2}}$, and a small number $\iota$. Calculate $\overline{\boldsymbol{G }}$  through one group $(\sigma _{1},\sigma _{2})$.
\STATE \textbf{For} $l=1,2,...,L$ 
\STATE \quad \quad \textbf{For} $t=1,2$
\STATE \quad \quad \quad \quad Substitute \eqref{ee42} into \eqref{ee41} and fix one of the $(\sigma _{1},\sigma _{2})$, we have
\STATE \quad \quad  \quad 
\begin{equation}
\label{ee43}
\sigma _{t}^{\circ }=   \mathop{\textup{arg\, min}}\limits_{\tau \in S_{\tau},\sigma _{1} \in S_{\sigma _{1}},\sigma _{2} \in S_{\sigma _{2}} }    \left [(\overline{\boldsymbol{G }}+\iota \mathbf{I})^{-1}\overline{\boldsymbol{h }}  \right ]^{T}\overline{\boldsymbol{G }} \left [(\overline{\boldsymbol{G }}+\iota \mathbf{I})^{-1}\overline{\boldsymbol{h }}  \right ]
\end{equation}
\STATE \quad \quad \textbf{End For}
\STATE \textbf{End For}

\STATE After the optimization of kernel width is completed, calculate the $\overline{\boldsymbol{\tau }} ^\circ$ by \eqref{ee42}. Then the fusion factor is obtained by $\tau ^\circ = \overline{\boldsymbol{\tau }}(1)$.
\end{algorithmic}
\label{AM2}
\end{algorithm}

\section{Performance Analysis}
\subsection{Numerical Stability Analysis}
Since the matrix $\Lambda $ is singular in \citep{8937723}, the MEE-KF has feeble stability, which often has large estimation deviation due to the interference of noise or outliers. Although a numerically stable MEE-KF is proposed in \citep{wang2021numerically}, such a stable local solution may blackuce the estimation accuracy. Fortunately, the proposed A-MEEF-UKF is still stable under the premise of optimal solution, as $\mathbf{\Phi} _{i}$ in \eqref{e26} is non-singular and symmetric positive definite.

\textit{Proof}: The elements in $\mathbf{\Phi} _{i}$ are defined as $\left \{ \mathbf{\Phi} _{i} \right \}_{jk}=c_{jk}$. Through \eqref{e24} and \eqref{e26}, the diagonal elements $ c_{kk} $ satisfy
\begin{equation}
\label{e38}
c_{kk}=\tau G_{\sigma _{1}}(e_{i,k})+(1-\tau)\sum_{j=1}^{N}\left (G_{\sigma _{2}}(e_{i,k}-e_{i,j})-1  \right )>0
\end{equation}
Excluding diagonal elements, the sum of the elements in $k$-th column is 
\begin{equation}
\label{e39}
\sum_{j=1,j\neq k}^{N}|c_{jk}|=(1-\tau )G_{\sigma _{2}}(e_{i,j}-e_{i,k})
\end{equation}
According to \eqref{e38} and \eqref{e39}, we have
\begin{equation}
\label{e40}
c_{kk}-\sum_{j=1,j\neq k}^{N}|c_{jk}|=\tau G_{\sigma _{1}}(e_{i,k})
\end{equation}
One can conclude that $\mathbf{\Phi} _{i}$ is a strictly diagonally dominant matrix, thus $\mathbf{\Phi} _{i}$ is non-singular.
$\mathbf{\Phi} _{i}$ is also positive definite because of $c_{kk}>0$. Therefore, the stability of the proposed algorithm is greatly improved compablack with MEE-UKF.

In addition,  the update of sigma vectors require $\mathbf{P}_{i|i-1}$ and $\mathbf{P}_{i|i}$ to be positive definite, otherwise the UKF estimator will be terminated.

\subsection{Convergence Analysis}
This section presents a sufficient condition for the convergence of fixed-point iteration in the proposed A-MEEF-UKF \citep{chen2015convergence,zhang2015convergence}. For simplicity, the kernel width of MEEF satisfies $\sigma_1=b\sigma_2$ with non-negative constant $b$. Through \eqref{e22}, the \eqref{e25} can be expressed as $\mathbf{x}_{i}=\mathbf{g}(\mathbf{x}_{i})=\mathbf{M}_{aa}\mathbf{N}_{az}$, where 
\begin{equation}
\label{e41}
\begin{aligned}
{{\bf{{\rm M}}}_{aa}} &= \tau \sum\limits_{k = 1}^N {{G_{{\sigma _1}}}\left( {{e_{i,k}}} \right){\bf{a}}_{i,k}^T{{\bf{a}}_{i,k}}} \\
 &+ (1 - \tau )\sum\limits_{j = 1}^N {\sum\limits_{k = 1}^N {{G_{{\sigma _2}}}\left( {{e_{i,j}} - {e_{i,k}}} \right)} } \left( {{\bf{a}}_{i,j}^T - {\bf{a}}_{i,k}^T} \right){{\bf{a}}_{i,j}}
\end{aligned}
\end{equation}
\begin{equation}
\label{e42}
\begin{aligned}
{{\bf{N}}_{az}} &= \tau \sum\limits_{k = 1}^N {{G_{{\sigma _1}}}\left( {{e_{i,k}}} \right){\bf{a}}_{i,k}^T{z_{i,k}}} \\
 &+ (1 - \tau )\sum\limits_{j = 1}^N {\sum\limits_{k = 1}^N {{G_{{\sigma _2}}}\left( {{e_{i,j}} - {e_{i,k}}} \right)} } \left( {{\bf{a}}_{i,j}^T - {\bf{a}}_{i,k}^T} \right){z_{i,j}}
\end{aligned}
\end{equation}
Note that the second item on the right hand in \eqref{e41} and \eqref{e42} is obviously more concise compablack with \citep{8937723,9646176,zhang2015convergence}. Therefore, the following convergence  analysis has less complexity.

\textit{Theorem 1}: If $\beta > \nu $,  and $\sigma _{2}>max(\sigma _{2}^{*},\sigma _{2}^{+})$, in which $\nu$ satisfies 
\begin{small}
\begin{equation}
\label{e45}
\begin{array}{l}
\nu  = \\
\frac{{\sqrt n \left( {\tau \sum\limits_{k = 1}^N {||{\bf{a}}_{i,k}^T|{|_1}|{z_{i,k}}} | + (1 - \tau )\sum\limits_{j = 1}^N {\sum\limits_{k = 1}^N {||{\bf{a}}_{i,j}^T - {\bf{a}}_{i,k}^T|{|_1}|{z_{i,j}}|} } } \right)}}{{{{\bf{\lambda }}_{\min }}\left[ {\tau \sum\limits_{k = 1}^N {{\bf{a}}_{i,k}^T{{\bf{a}}_{i,k}}}  + (1 - \tau )\sum\limits_{j = 1}^N {\sum\limits_{k = 1}^N {\left( {{\bf{a}}_{i,j}^T - {\bf{a}}_{i,k}^T} \right){{\bf{a}}_{i,j}}} } } \right]}}
\end{array}
\end{equation}
\end{small}
and $\sigma _{2}^{*}$ is the solution  of $\phi (\sigma _{2})=\beta $, with
\begin{small}
\begin{equation}
\label{e46}
\begin{array}{l}
\phi ({\sigma _2}) = \\
\frac{{\sqrt n \left( {\tau \sum\limits_{k = 1}^N {||{\bf{a}}_{i,k}^T|{|_1}|{z_{i,k}}} | + (1 - \tau )\sum\limits_{j = 1}^N {\sum\limits_{k = 1}^N {||{\bf{a}}_{i,j}^T - {\bf{a}}_{i,k}^T|{|_1}} } |{z_{i,j}}|} \right)}}{{{{\bf{\lambda }}_{\min }}\left[ {\tau \sum\limits_{k = 1}^N {{G_{{\sigma _1}}}\left( {{\eta _1}} \right){\bf{a}}_{i,k}^T{{\bf{a}}_{i,k}}}  + (1 - \tau )\sum\limits_{j = 1}^N {\sum\limits_{k = 1}^N {{G_{{\sigma _2}}}\left( {{\eta _2}} \right)\left( {{\bf{a}}_{i,j}^T - {\bf{a}}_{i,k}^T} \right){{\bf{a}}_{i,j}}} } } \right]}}
\end{array}
\end{equation}
\end{small}
and $\sigma _{2}^{+}$ is the solution  of $\varphi  (\sigma _{2})=\alpha (0<\alpha<1)$, with
\begin{small}
\begin{equation}
\label{e47}
\begin{array}{l}
\varphi ({\sigma _2}) = \\
\frac{{\tau \sqrt n \sum\limits_{k = 1}^N {{\eta _1}||{{\bf{a}}_{i,k}}|{|_1}\left( {\beta ||{\bf{a}}_{i,k}^T{{\bf{a}}_{i,k}}|{|_1} + ||{\bf{a}}_{i,k}^T{z_{i,k}}|{|_1}} \right)} }}{{b^2\sigma _2^2{{\bf{\lambda }}_{\min }}\left[ {\sum\limits_{k = 1}^N {{G_{{\sigma _1}}}\left( {{\eta _1}} \right){\bf{a}}_{i,k}^T{{\bf{a}}_{i,k}}} } \right]}} + \frac{{(1 - \tau )\sqrt n }}{{\sigma _2^2}} \times \\
\frac{{\sum\limits_{j = 1}^N {\sum\limits_{k = 1}^N {{\eta _2}||{\bf{a}}_{i,j} - {\bf{a}}_{i,k}|{|_1}\left( {\beta {{\left\| {\left( {{\bf{a}}_{i,j}^T - {\bf{a}}_{i,k}^T} \right){{\bf{a}}_{i,j}}} \right\|}_1} + {{\left\| {\left( {{\bf{a}}_{i,j}^T - {\bf{a}}_{i,k}^T} \right){z_{i,j}}} \right\|}_1}} \right)} } }}{{{{\bf{\lambda }}_{\min }}\left[ {\sum\limits_{j = 1}^N {\sum\limits_{k = 1}^N {{G_{{\sigma _2}}}\left( {{\eta _2}} \right)\left( {{\bf{a}}_{i,j}^T - {\bf{a}}_{i,k}^T} \right){{\bf{a}}_{i,j}}} } } \right]}}
\end{array}
\end{equation}
\end{small}
where $\eta _{1}=\beta ||\mathbf{a}_{i,k}||_{1}+|z_{i,k}|$, $\eta _{2}=\beta ||\mathbf{a}_{i,j}-\mathbf{a}_{i,k}||_{1}+|z_{i,j}-z_{i,k}|$, and $\mathbf{\lambda} _{min}[\cdot ]$ denotes the minimum eigenvalue of matrix. 
Then it holds that $||\mathbf{g}(\mathbf{x}_{i})||_{1}\leq \beta $, and $||\nabla_{\mathbf{x}_{i}}\mathbf{g}(\mathbf{x}_{i})||_{1}\leq \alpha $ for all $\mathbf{x}_{i}\in \left \{ \mathbf{x}_{i}\in \mathbb{R}^{n}:|| \mathbf{x}_{i}||_{1}\leq \beta  \right \}$, 
where the gradient of $\mathbf{g}(\mathbf{x}_{i})$ is given by
\begin{equation}
\label{e43}
{\nabla _{{{\bf{x}}_i}}}{\bf{g}}\left( {{{\bf{x}}_i}} \right) = \left[ {\frac{{\partial {\bf{g}}\left( {{{\bf{x}}_i}} \right)}}{{\partial {x_{i,1}}}},\frac{{\partial {\bf{g}}\left( {{{\bf{x}}_i}} \right)}}{{\partial {x_{i,2}}}}, \cdots ,\frac{{\partial {\bf{g}}\left( {{{\bf{x}}_i}} \right)}}{{\partial {x_{i,n}}}}} \right]
\end{equation}
with 
\begin{small}
\begin{equation}
\label{e44}
\begin{array}{l}
\frac{{\partial {\bf{g}}\left( {{{\bf{x}}_i}} \right)}}{{\partial {x_{i,f}}}} = \\
{\bf{{\rm M}}}_{aa}^{ - 1}\left( {\sum\limits_{k = 1}^N {{\mu _1}{\bf{a}}_{i,k}^T{{\bf{a}}_{i,k}}}  + } \right.\left. {\sum\limits_{j = 1}^N {\sum\limits_{k = 1}^N {{\mu _2}\left( {{\bf{a}}_{i,j}^T - {\bf{a}}_{i,k}^T} \right){{\bf{a}}_{i,j}}} } } \right){\bf{g}}\left( {{{\bf{x}}_i}} \right)\\
{\kern 1pt} {\kern 1pt} {\kern 1pt}  + {\bf{{\rm M}}}_{aa}^{ - 1}\left( {\sum\limits_{k = 1}^N {{\mu _1}{\bf{a}}_{i,k}^T{z_{i,k}}}  + } \right.\left. {\sum\limits_{j = 1}^N {\sum\limits_{k = 1}^N {{\mu _2}\left( {{\bf{a}}_{i,j}^T - {\bf{a}}_{i,k}^T} \right){z_{i,j}}} } } \right)
\end{array}
\end{equation}
\end{small}
where ${\mu _1} = \frac{\tau }{{b^2\sigma _2^2}}{e_{i,k}}a_{i,k}^f{G_{{\sigma _1}}}\left( {{e_{i,k}}} \right)$, ${\mu _2} = \frac{{1 - \tau }}{{\sigma _2^2}}({e_{i,j}} - {e_{i,k}})(a_{i,j}^f - a_{i,k}^f){G_{{\sigma _2}}}\left( {{e_{i,j}} - {e_{i,k}}} \right)$.   
$x_{i,f}$ and $a_{i,k}^f$ denote $f$-th elements of $\mathbf{x}_{i}$ and $\mathbf{a}_{i,k}$, respectively. 
 
By the Theorem 1 and Banach Fixed-Point Theorem, given an initial iterative value  $||\mathbf{x}_{i}^0||_1<\beta$,  proper kernel width  $\sigma _{2}>max(\sigma _{2}^{*},\sigma _{2}^{+})$, the fixed-point iteration in the A-MEEF-UKF can converge to a unique $\mathbf{x}_{i}$.

\section{Simulation Results}
This section provides two nonlinear examples to verify the excellent performance of the proposed A-MEEF-UKF  in different noise environments. The standard UKF, \textcolor{black}{the Sage-Husa adaptive UKF (AUKF)  \citep{zhao2010adaptive}}, the MCC-UKF \citep{liu2017maximum}, the  MEE-UKF \citep{dang2020robust}, the robust MEE-UKF (R-MEE-UKF) \citep{wang2021numerically}, \textcolor{black}{ and the MEEF-CKF \citep{9646176}} are used as the control algorithms. In order to avoid contingency, all results are obtained through $M = 100$ independent Monte Carlo runs. And the root mean square error ($RMSE$) is defined by 
\begin{equation}
\label{e48}
RMSE_{i}=\sqrt{\frac{1}{M}\sum_{l=1}^{M}||\mathbf{\hat{x}}_{i|i}(l)-\mathbf{x}_{i}(l)||_{2}^{2}}
\end{equation}
where $\mathbf{\hat{x}}_{i|i}(l)$ and $\mathbf{x}_{i}(l)$ respectively represent estimated state and true state at $l$-th Monte Carlo experiment.

\subsection{Example 1 }
\textcolor{black}{This example will illustrate the stability of A-MEEF-UKF in the univariate non-stationary growth model, which is a representative of strongly nonlinear systems \citep{liu2017maximum} .  The state and measurement equations are given by } 
\begin{equation}
\label{e49}
x_{i}=0.5x_{i-1}+\frac {25x_{i-1}}{1+x_{i-1}^{2}}+8cos(1.2x_{i-1})+q_{i-1}
\end{equation}
\begin{equation}
\label{e50}
y_{i}=\frac{x_{i}^{2}}{20}+r_{i}
\end{equation}
The process noise $q_{i-1}$ and measurement noise $r_{i}$ are known, obeying Gaussian distribution. The initial true state, estimated state, and covariance matrix are $x_{0}=0$,  $\hat{x}_{0|0}=1$, and  $P_{0|0}=10$, respectively. The unscented transformation parameters are set to $\bar{\alpha }=1$, $\bar{\beta  }=0$, and $\kappa =3-n$ \citep{kulikov2021ito}.  The iteration threshold is set to $\gamma =10^{-6}$. \textcolor{black}{Generally, the forgetting factor $b$ in $d_i$ can be taken as a number close to 1, here we set $b=0.998$.}
 
\textcolor{black}{For the small noise  $q_{i-1}\sim N(0,0.10^{-6})$ and $r_{i}\sim N(0,0.01)$, where the 0 is  mean and the 0.01 is variance}, Table \ref{tab:table1} shows the average $RMSE$ of state in this scenario, it can be seen that these algorithms have the similar performance  under small Gaussian noise. Because the MEE-UKF has poor stability, it can not achieve accurate estimation results in this noise environment even if a large kernel width is employed.

\textcolor{black}{To further illustrate the improvement of the MEEF criterion in stability compablack with the MEE, we employ larger noise interference in this case: $q_{i-1}\sim N(0,1)$ and $r_{i}\sim N(0,100)$. The average $RMSE$ of state in Table \ref{tab:table2} displays that both of the A-MEEF-UKF and MEEF-CKF can maintain stability, which is mainly attributed to the fact that matrix $\mathbf{\Phi} _{i}$ is positive definite in the MEEF criterion.}
\begin{table}[!t]
\caption{ \textcolor{black}{Average $RMSE$ of state under small Gaussian noise for example 1.}\label{tab:table1} }
\centering
\begin{tabular}{c c}
\hline
Algorithms & Average $RMSE$ \\
\hline
UKF & 15.3869 \\
AUKF & 14.1192 \\
MCC-UKF & 22.5050\\
MEE-UKF & 79.1097 \\
R-MEE-UKF & 28.2602 \\
MEEF-CKF & 15.4524 \\
A-MEEF-UKF & 16.4409\\
\hline
\end{tabular}
\end{table}

\begin{table}[!t]
\caption{ \textcolor{black}{Average $RMSE$ of state under large Gaussian noise for example 1.\label{tab:table2} }}
\centering
\begin{tabular}{c c}
\hline
Algorithms & Average $RMSE$ \\
\hline
UKF & 12.5941 \\
AUKF & 12.6728 \\
MCC-UKF & 12.8053\\
MEE-UKF & $\infty $ \\
R-MEE-UKF & 13.4503 \\
MEEF-CKF & 13.3505 \\
A-MEEF-UKF & 13.6870\\
\hline
\end{tabular}
\end{table}

\subsection{Example 2 }
Consider an example of vehicle navigation \citep{8937723}. \textcolor{black}{The state equation}  is given by 
\textcolor{black}{
\begin{equation}
\label{ee51}
[\dot{x}_{1},\dot{x}_{2},\dot{x}_{3},\dot{x}_{4}]^{T} =  [x_{3},x_{4},0,0]^{T} + \mathbf{q}_{i}
\end{equation}
where the state  $\dot{x}_{1}$,$\dot{x}_{2}$,$\dot{x}_{3}$,and $\dot{x}_{4}$ }are the north position, east position, north velocity, and east velocity. \textcolor{black}{Taking the sampling interval $\Delta T = 0.2s$}, \eqref{ee51} can be blackuced to the following discrete form
\begin{equation}
\label{e51}
{{\bf{x}}_i} = \left[ {\begin{array}{*{20}{c}}
1&0& \Delta T &0\\
0&1&0& \Delta T \\
0&0&1&0\\
0&0&0&1
\end{array}} \right]{{\bf{x}}_{i - 1}} + {{\bf{q}}_{i - 1}}
\end{equation}
where the discrete state is $\mathbf{x}_{i}=[x_{i,1},x_{i,2},x_{i,3},x_{i,4}]^{T}$.
In addition, we adopt the following measurement equation:
\begin{equation}
\label{e52}
{{\bf{y}}_i} = \left[ {\begin{array}{*{20}{c}}
{ - {x_{i,1}} - {x_{i,3}}}\\
{ - {x_{i,2}} - {x_{i,4}}}\\
{\sqrt {x_{i,1}^2 + x_{i,2}^2} }\\
{\arctan \sqrt {\frac{{{x_{i,2}} - \bar x}}{{{x_{i,1}} - \bar y}}} }
\end{array}} \right] + {{\bf{r}}_i}
\end{equation}
where the position of the measuring instrument is $\bar{x}=-100$ and $\bar{y}=-100$.

The initial true state, estimated state, and covariance matrix are $x_{0}=[0,0,5,10]^T$,  $\hat{x}_{0|0}=[1,1,4,8]^T$, and  $P_{0|0}=\textup{diag}[10,10,50,100]^T$, respectively. These parameters are also set to $\bar{\alpha }=1$, $\bar{\beta  }=0$,  $\kappa =3-n$,  $\gamma =10^{-6}$, and $b = 0.998$. \textcolor{black}{The kernel widths of MCC-UKF, MEE-UKF, and R-MEE-UKF are all generated by training, and the free parameter of MEEF-CKF and A-MEEF-UKF are online calculated by Algorithm 2}. The process noise obeys Gaussian distribution $\mathbf{q}_{i-1}\sim N(0,0.001)$. Consider the following \textcolor{black}{four} cases of measurement  noise. \textcolor{black}{ Case (a): Gaussian noise, where $\mathbf{r}_i\sim N(0,10)$.} Case (b): Gaussian noise with random outliers (heavy-tailed noise), where $\mathbf{r}_i\sim 0.99N(0,0.001)+0.01N(0,1000)$. Case (c): Bimodal Gaussian noise with random outliers, where $\mathbf{r}_i\sim 0.49N(-0.1,0.001)+0.49N(0.1,0.001)+0.02N(0,1000)$. Case (d): Asymmetric noise, where $\mathbf{r}_i \sim  0.99N(-0.1,0.001)+0.01N(100,1000)$. Under different noise environments, the $RMSE$ of several algorithms is studied below:

\subsubsection{Known Noise}
\begin{figure}[!t]
\centering
\includegraphics[width=4.5in]{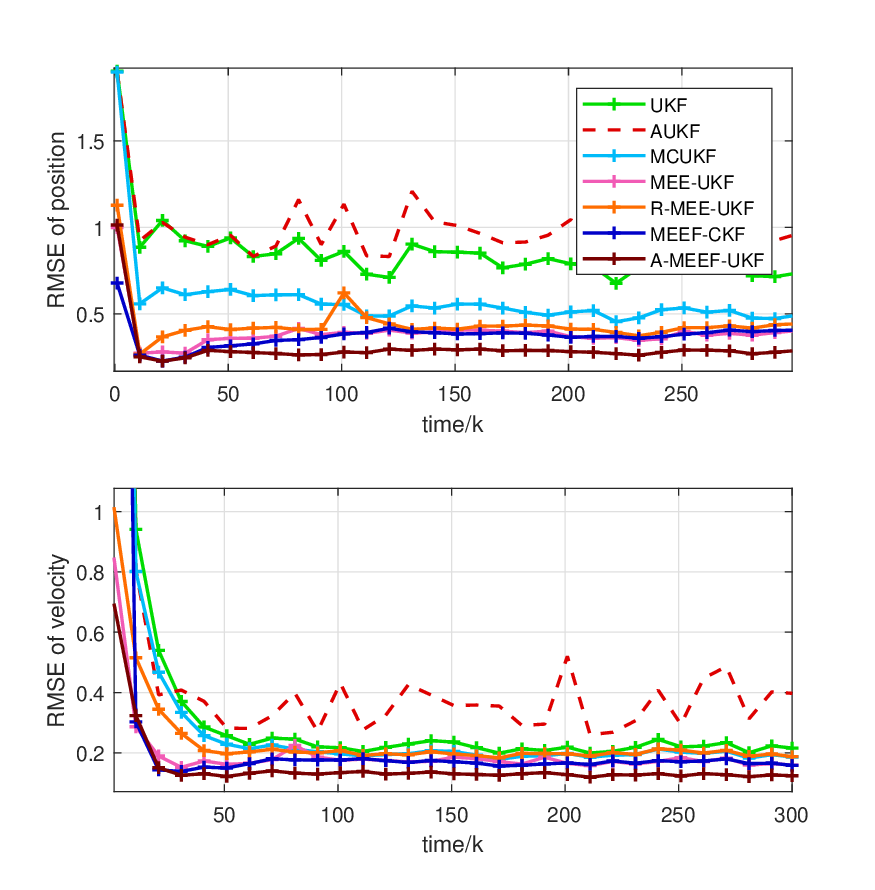}
\caption{\textcolor{black}{$RMSE$ of position and velocity in several algorithms for Case (b) of example 2 when noise distribution is assumed to be known.}}
\label{fig_1}
\end{figure}

\begin{figure}[!t]
\centering
\includegraphics[width=4.5in]{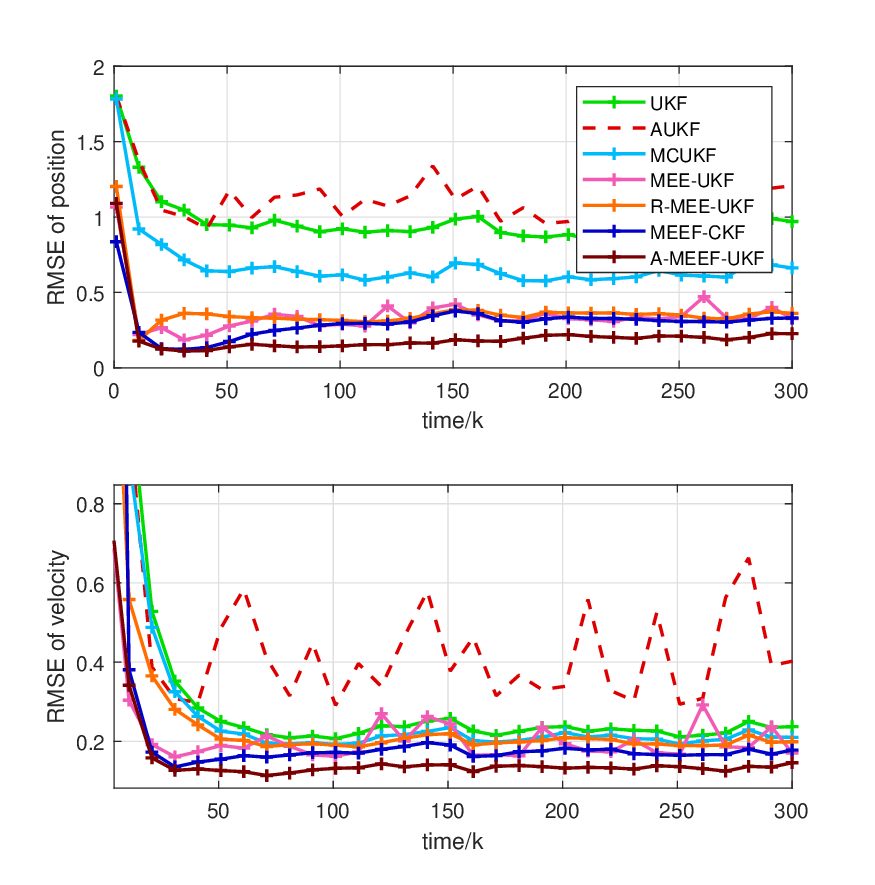}
\caption{\textcolor{black}{$RMSE$ of position and velocity in several algorithms for Case (c) of example 2 when noise distribution is assumed to be known.}}
\label{fig_2}
\end{figure}

\begin{figure}[!t]
\centering
\includegraphics[width=4.5in]{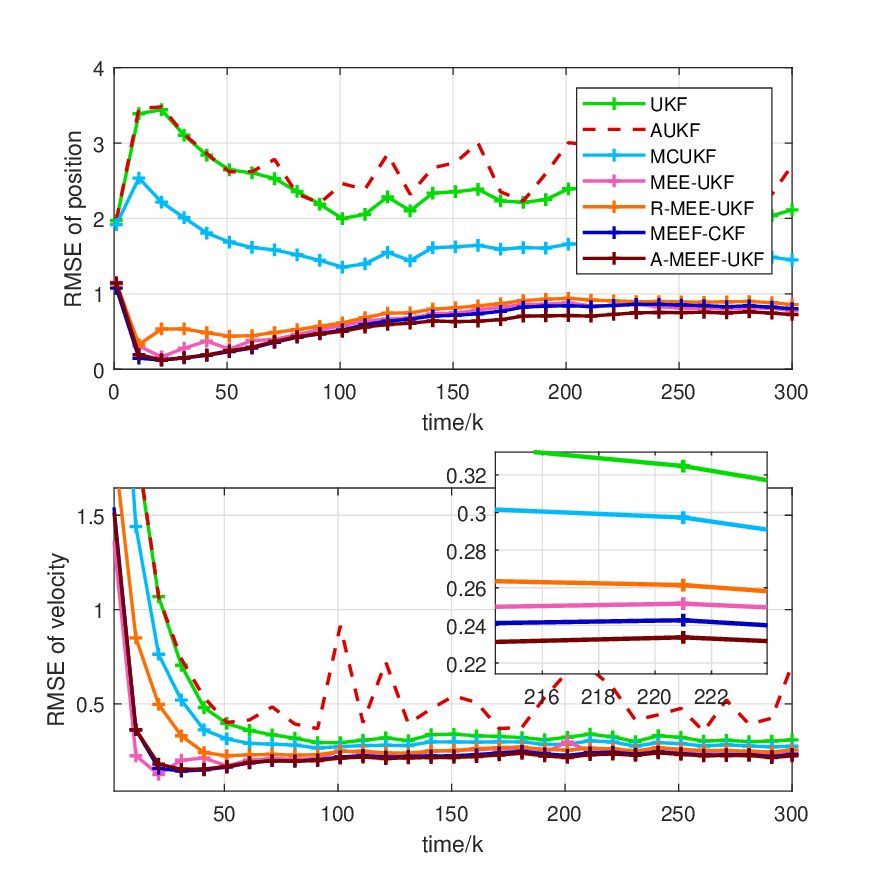}
\caption{\textcolor{black}{$RMSE$ of position and velocity in several algorithms for Case (d) of example 2 when noise distribution is assumed to be known.}}
\label{fig_3}
\end{figure}

In this scenario, the noise distribution is assumed to be known,  \textcolor{black}{namely $\mathbf{\hat{Q}}_{0}=\mathbf{Q}_{0}$ and $\mathbf{\hat{R}}_{1}=\mathbf{R}_{1}$.}
Table \ref{tab:table3} and Table \ref{tab:table4} show that  the average $RMSE$ of position and velocity in several algorithms, where the $RMSE$ of position is calculated by $x_{i,1}$ and  $x_{i,2}$, the $RMSE$ of velocity is determined by $x_{i,3}$ and  $x_{i,4}$. One can conclude that the proposed algorithm has the best performance when measurements suffer from complex non-Gaussian noises. \textcolor{black}{In addition, the $RMSE$ of several algorithms at each time is given in Fig.\ref{fig_1}, Fig.\ref{fig_2} and  Fig.\ref{fig_3}, where the k stands for sampling time point. The A-MEEF-UKF has superior estimation performance by means of the MEEF criterion. Specifically, the error entropy can well simulate the error distribution, and the correntropy  can resist impulse noise or outliers.}
\begin{table}[!t]
\caption{\textcolor{black}{Average $RMSE$ of position in several algorithms for example 2 when noise distribution is assumed to be known.}
\label{tab:table3}}
\centering
\begin{tabular}{c c c c c}
\hline
Algorithms & Case (a) & Case (b) & Case (c) & Case (d) \\
\hline
UKF         &0.9072&0.8277&0.9500&2.4000\\
AUKF        &1.0450&0.9649&1.0873&2.6638\\
MCC-UKF     &0.9074&0.5551&0.6594&1.6505\\
MEE-UKF     &1.1968&0.3780&0.3325&0.6623\\
R-MEE-UKF   &1.0378&0.4245&0.3499&0.6179\\
MEEF-CKF    &1.0430&0.3746&0.2988&0.5628\\
A-MEEF-UKF  &1.0977&\textbf{0.2862}&\textbf{0.1868}&0.6388\\
\hline
\end{tabular}
\end{table}

\begin{table}[!t]
\caption{\textcolor{black}{Average $RMSE$ of velocity in several algorithms for example 2 when noise distribution is assumed to be known.}
\label{tab:table4}}
\centering
\begin{tabular}{c c c c c}
\hline
Algorithms & Case (a) & Case (b) & Case (c) & Case (d) \\
\hline
UKF         &0.3279&0.3058&0.3306&0.4780\\
AUKF        &0.4637&0.4269&0.4626&0.6419\\
MCC-UKF     &0.3279&0.2753&0.3052&0.4123\\
MEE-UKF     &0.7998&0.1898&0.2123&0.2623\\
R-MEE-UKF   &0.4142&0.2283&0.2379&0.3062\\
MEEF-CKF    &0.3461&0.2158&0.2289&0.2379\\
A-MEEF-UKF  &0.5376&\textbf{0.1466}&\textbf{0.1531}&\textbf{0.2331}\\
\hline
\end{tabular}
\end{table}

\subsubsection{Unknown Noise}
\begin{figure}[!t]
\centering
\includegraphics[width=4.5in]{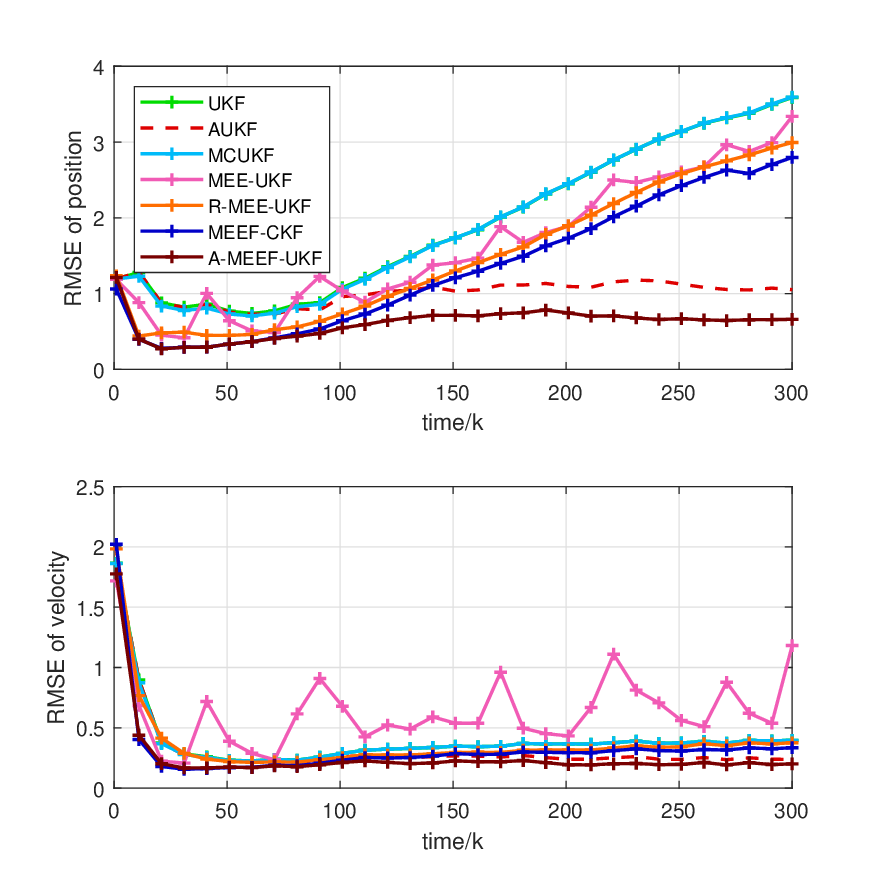}
\caption{\textcolor{black}{$RMSE$ of position and velocity in several algorithms for Case (b) of example 2 when noise distribution is unknown.}}
\label{fig_4}
\end{figure}

\begin{figure}[!t]
\centering
\includegraphics[width=4.5in]{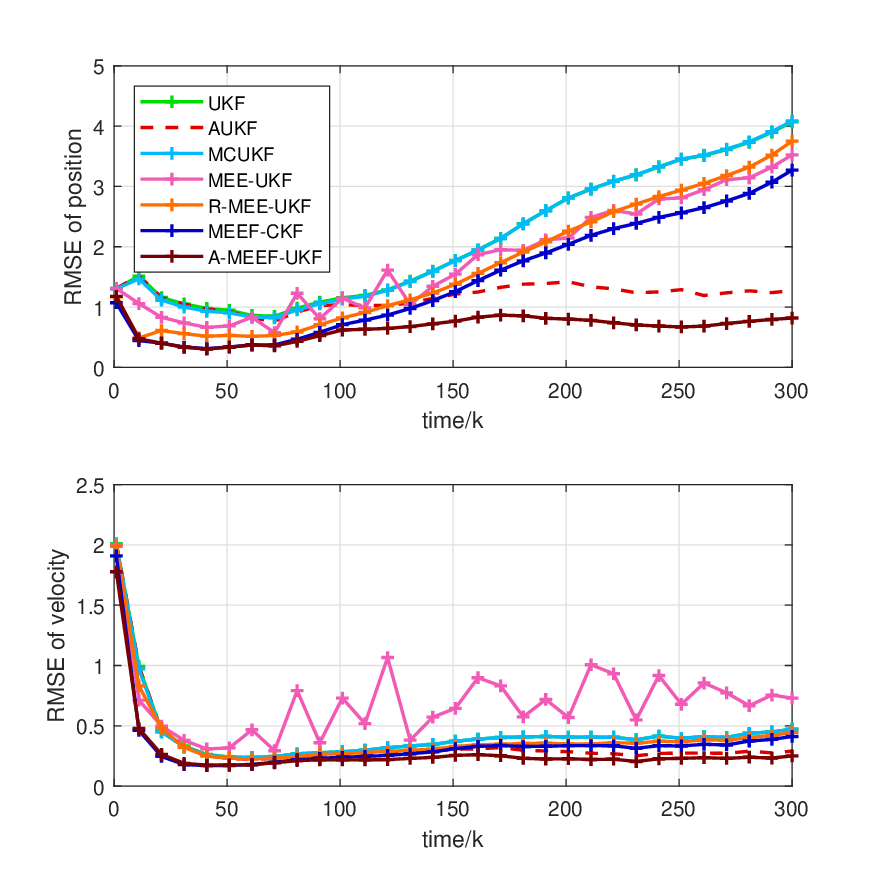}
\caption{\textcolor{black}{$RMSE$ of position and velocity in several algorithms for Case (c) of example 2 when noise distribution is unknown.}}
\label{fig_5}
\end{figure}

\begin{figure}[!t]
\centering
\includegraphics[width=4.5in]{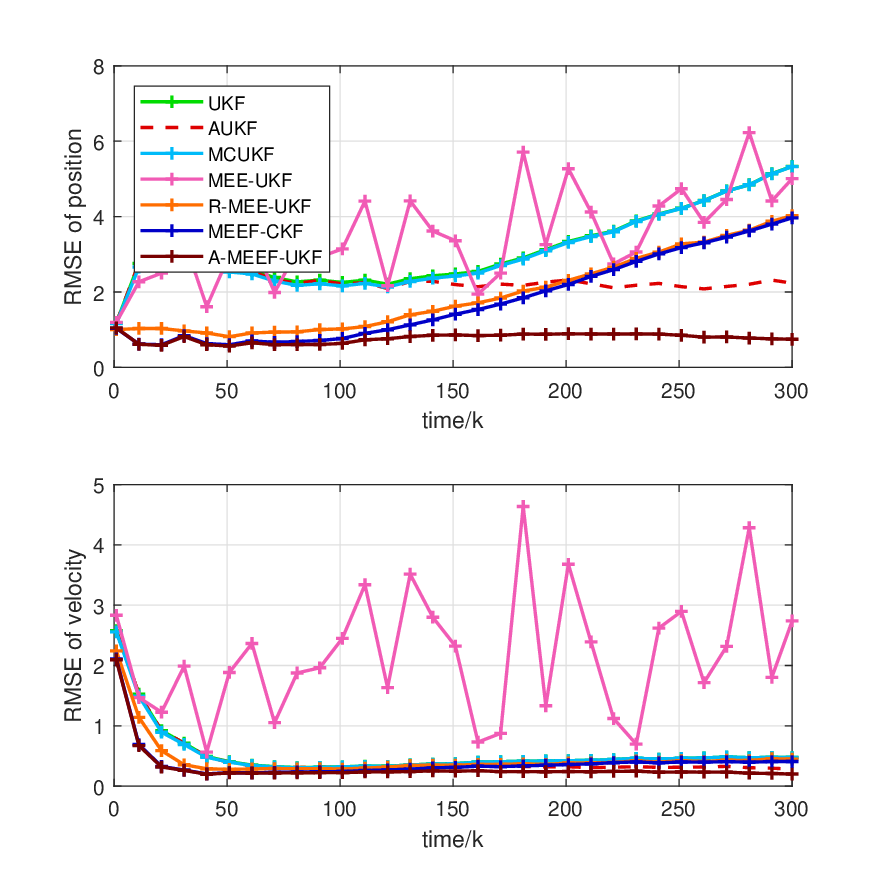}
\caption{\textcolor{black}{$RMSE$ of position and velocity in several algorithms for Case (d) of example 2 when noise distribution is unknown.}}
\label{fig_6}
\end{figure}

\textcolor{black}{In practice, the distribution characteristics of noise may be unknown, and we can not accurately obtain the noise covariance matrix.} Considering that the noise distribution is unknown in this scenario, the initial estimated noise covariance matrices are $\mathbf{\hat{Q}}_{0}=0.01\mathbf{Q}_{0}$ and $\mathbf{\hat{R}}_{1}=10\mathbf{R}_{1}$. Table \ref{tab:table5} and Table \ref{tab:table6} show the average $RMSE$ of position and velocity, respectively. \textcolor{black}{Fig. \ref{fig_4}, Fig. \ref{fig_5} and Fig. \ref{fig_6} display the fluctuation  of $RMSE$ over time. Obviously, by adopting adaptive noise covariance matrix, the AUKF and the proposed A-MEEF-UKF show excellent results. As time goes on, the estimates of other algorithms gradually diverge.}

\begin{table}[!t]
\caption{\textcolor{black}{Average $RMSE$ of position in several algorithms for example 2 when noise distribution is unknown.}
\label{tab:table5}}
\centering
\begin{tabular}{c c c c c}
\hline
Algorithms & Case (a) & Case (b) & Case (c) & Case (d) \\
\hline
UKF         &2.0034&1.9029&2.0957&3.1666\\
AUKF        &1.2160&1.0055&1.1631&2.3476\\
MCC-UKF     &2.0041&1.8913&2.0834&3.1098\\
MEE-UKF     &1.8031&1.5910&1.7936&3.4361\\
R-MEE-UKF   &1.6810&1.4466&1.6566&1.9051\\
MEEF-CKF    &1.5945&1.3127&1.4397&1.7413\\
A-MEEF-UKF  &\textbf{1.0884}&\textbf{0.5888}&\textbf{0.6361}&\textbf{0.7660}\\
\hline
\end{tabular}
\end{table}

\begin{table}[!t]
\caption{\textcolor{black}{Average $RMSE$ of velocity in several algorithms for example 2 when noise distribution is unknown.}
\label{tab:table6}}
\centering
\begin{tabular}{c c c c c}
\hline
Algorithms & Case (a) & Case (b) & Case (c) & Case (d) \\
\hline
UKF         &0.3948&0.3747&0.4042&0.5062\\
AUKF        &0.3366&0.3000&0.3283&0.4395\\
MCC-UKF     &0.3948&0.3730&0.4027&0.5015\\
MEE-UKF     &0.8453&0.5850&0.7010&2.0317\\
R-MEE-UKF   &0.3937&0.3399&0.3706&0.4212\\
MEEF-CKF    &0.3486&0.2864&0.3117&0.3564\\
A-MEEF-UKF  &\textbf{0.315}8&\textbf{0.2273}&\textbf{0.2497}&\textbf{0.2772}\\
\hline
\end{tabular}
\end{table}

\subsubsection{Computational burden}
\textcolor{black}{In this section, we give the computational burden of these algorithms. Table \ref{tab:table7} shows the computational time of several algorithms for performing an estimate. It can be seen that the execution time of the proposed A-MEEF-UKF is completely within the acceptable range, because it is far below  the sampling time $\Delta T = 0.2s$. In fact, the time consumption of A-MEEF-UKF is mainly reflected in the online selection of free parameters, which is why we execute Algorithm 2 in a finite set.}

\begin{table}[!t]
\caption{\textcolor{black}{Single step operation time for example 2.}
\label{tab:table7}}
\centering
\begin{tabular}{c c }
\hline
Algorithms & Single step operation time  \\
\hline
UKF         & 0.000159 s   \\
AUKF        & 0.000244 s   \\
MCC-UKF     & 0.000282 s   \\
MEE-UKF     & 0.000324 s   \\
R-MEE-UKF   & 0.000328 s   \\
MEEF-CKF    & 0.001914 s  \\
A-MEEF-UKF  & 0.001301 s  \\
\hline
\end{tabular}
\end{table}

\section{Conclusion}
This paper develops a novel robust unscented Kalman filter. It has the following advantages: 1) Based on minimum error entropy with fiducial points criterion, the novel algorithm has the common advantages of correntropy and error entropy, which exhibits superior robustness to complex non-Gaussian noises and outliers. 2) Compablack with the traditional error entropy unscented Kalman filter, the novel algorithm has better numerical stability. 3) The improved Sage-Husa estimation can timely and accurately correct the covariance matrix of process noise and measurement noise, effectively addressing the problem of state estimation in noisy uncertain systems. Finally, the robustness and accuracy of the proposed algorithm are reliably verified by Monte Carlo experiments under different noise environments.

\section*{Acknowledgements}
This work was partially supported by National Natural Science Foundation of China (grant: 62171388, 61871461, 61571374), and  Fundamental Research Funds for the Central Universities (grant: 2682021ZTPY091).

\bibliographystyle{elsarticle-num-names} 
\bibliography{cas-refs}

\begin{thebibliography}{31}
\expandafter\ifx\csname natexlab\endcsname\relax\def\natexlab#1{#1}\fi
\providecommand{\url}[1]{\texttt{#1}}
\providecommand{\href}[2]{#2}
\providecommand{\path}[1]{#1}
\providecommand{\DOIprefix}{doi:}
\providecommand{\ArXivprefix}{arXiv:}
\providecommand{\URLprefix}{URL: }
\providecommand{\Pubmedprefix}{pmid:}
\providecommand{\doi}[1]{\href{http://dx.doi.org/#1}{\path{#1}}}
\providecommand{\Pubmed}[1]{\href{pmid:#1}{\path{#1}}}
\providecommand{\bibinfo}[2]{#2}
\ifx\xfnm\relax \def\xfnm[#1]{\unskip,\space#1}\fi
\bibitem[{Liu et~al.(2007)Liu, Pokharel, and Principe}]{liu2007correntropy}
\bibinfo{author}{W.~Liu}, \bibinfo{author}{P.~P. Pokharel}, \bibinfo{author}{J.~C. Principe},
\newblock \bibinfo{title}{Correntropy: Properties and applications in non-gaussian signal processing},
\newblock \bibinfo{journal}{IEEE Transactions on signal processing} \bibinfo{volume}{55} (\bibinfo{year}{2007}) \bibinfo{pages}{5286--5298}.
\bibitem[{Chen et~al.(2018)Chen, Wang, Lu, Wang, Cao, and Qin}]{chen2018mixture}
\bibinfo{author}{B.~Chen}, \bibinfo{author}{X.~Wang}, \bibinfo{author}{N.~Lu}, \bibinfo{author}{S.~Wang}, \bibinfo{author}{J.~Cao}, \bibinfo{author}{J.~Qin},
\newblock \bibinfo{title}{Mixture correntropy for robust learning},
\newblock \bibinfo{journal}{Pattern Recognition} \bibinfo{volume}{79} (\bibinfo{year}{2018}) \bibinfo{pages}{318--327}.
\bibitem[{Chen et~al.(2021)Chen, Xie, Wang, Yuan, Ren, and Qin}]{chen2021multikernel}
\bibinfo{author}{B.~Chen}, \bibinfo{author}{Y.~Xie}, \bibinfo{author}{X.~Wang}, \bibinfo{author}{Z.~Yuan}, \bibinfo{author}{P.~Ren}, \bibinfo{author}{J.~Qin},
\newblock \bibinfo{title}{Multikernel correntropy for robust learning},
\newblock \bibinfo{journal}{IEEE Transactions on Cybernetics}  (\bibinfo{year}{2021}). \bibinfo{note}{Doi{: 10.1109/TCYB.2021.3110732}}.
\bibitem[{Chen et~al.(2014)Chen, Xing, Liang, Zheng, and Principe}]{chen2014steady}
\bibinfo{author}{B.~Chen}, \bibinfo{author}{L.~Xing}, \bibinfo{author}{J.~Liang}, \bibinfo{author}{N.~Zheng}, \bibinfo{author}{J.~C. Principe},
\newblock \bibinfo{title}{Steady-state mean-square error analysis for adaptive filtering under the maximum correntropy criterion},
\newblock \bibinfo{journal}{IEEE signal processing letters} \bibinfo{volume}{21} (\bibinfo{year}{2014}) \bibinfo{pages}{880--884}.
\bibitem[{Chen et~al.(2016)Chen, Xing, Zhao, Zheng, Pr{\i} et~al.}]{chen2016generalized}
\bibinfo{author}{B.~Chen}, \bibinfo{author}{L.~Xing}, \bibinfo{author}{H.~Zhao}, \bibinfo{author}{N.~Zheng}, \bibinfo{author}{J.~C. Pr{\i}}, et~al.,
\newblock \bibinfo{title}{Generalized correntropy for robust adaptive filtering},
\newblock \bibinfo{journal}{IEEE Transactions on Signal Processing} \bibinfo{volume}{64} (\bibinfo{year}{2016}) \bibinfo{pages}{3376--3387}.
\bibitem[{Chen et~al.(2017)Chen, Liu, Zhao, and Principe}]{chen2017maximum}
\bibinfo{author}{B.~Chen}, \bibinfo{author}{X.~Liu}, \bibinfo{author}{H.~Zhao}, \bibinfo{author}{J.~C. Principe},
\newblock \bibinfo{title}{Maximum correntropy kalman filter},
\newblock \bibinfo{journal}{Automatica} \bibinfo{volume}{76} (\bibinfo{year}{2017}) \bibinfo{pages}{70--77}.
\bibitem[{Liu et~al.(2021)Liu, Ren, Lyu, Jiang, Ren, and Chen}]{8736038}
\bibinfo{author}{X.~Liu}, \bibinfo{author}{Z.~Ren}, \bibinfo{author}{H.~Lyu}, \bibinfo{author}{Z.~Jiang}, \bibinfo{author}{P.~Ren}, \bibinfo{author}{B.~Chen},
\newblock \bibinfo{title}{Linear and nonlinear regression-based maximum correntropy extended kalman filtering},
\newblock \bibinfo{journal}{IEEE Transactions on Systems, Man, and Cybernetics: Systems} \bibinfo{volume}{51} (\bibinfo{year}{2021}) \bibinfo{pages}{3093--3102}.
\bibitem[{Liu et~al.(2017)Liu, Chen, Xu, Wu, and Honeine}]{liu2017maximum}
\bibinfo{author}{X.~Liu}, \bibinfo{author}{B.~Chen}, \bibinfo{author}{B.~Xu}, \bibinfo{author}{Z.~Wu}, \bibinfo{author}{P.~Honeine},
\newblock \bibinfo{title}{Maximum correntropy unscented filter},
\newblock \bibinfo{journal}{International Journal of Systems Science} \bibinfo{volume}{48} (\bibinfo{year}{2017}) \bibinfo{pages}{1607--1615}.
\bibitem[{Wang et~al.(2017)Wang, Li, and Zhang}]{wang2017maximum}
\bibinfo{author}{G.~Wang}, \bibinfo{author}{N.~Li}, \bibinfo{author}{Y.~Zhang},
\newblock \bibinfo{title}{Maximum correntropy unscented kalman and information filters for non-gaussian measurement noise},
\newblock \bibinfo{journal}{Journal of the Franklin Institute} \bibinfo{volume}{354} (\bibinfo{year}{2017}) \bibinfo{pages}{8659--8677}.
\bibitem[{Ma et~al.(2019)Ma, Qiu, Liu, Xiao, Duan, and Chen}]{ma2019unscented}
\bibinfo{author}{W.~Ma}, \bibinfo{author}{J.~Qiu}, \bibinfo{author}{X.~Liu}, \bibinfo{author}{G.~Xiao}, \bibinfo{author}{J.~Duan}, \bibinfo{author}{B.~Chen},
\newblock \bibinfo{title}{Unscented kalman filter with generalized correntropy loss for robust power system forecasting-aided state estimation},
\newblock \bibinfo{journal}{IEEE Transactions on Industrial Informatics} \bibinfo{volume}{15} (\bibinfo{year}{2019}) \bibinfo{pages}{6091--6100}.
\bibitem[{Liu et~al.(2018)Liu, Qu, Zhao, and Yue}]{liu2018maximum}
\bibinfo{author}{X.~Liu}, \bibinfo{author}{H.~Qu}, \bibinfo{author}{J.~Zhao}, \bibinfo{author}{P.~Yue},
\newblock \bibinfo{title}{Maximum correntropy square-root cubature kalman filter with application to sins/gps integrated systems},
\newblock \bibinfo{journal}{ISA transactions} \bibinfo{volume}{80} (\bibinfo{year}{2018}) \bibinfo{pages}{195--202}.
\bibitem[{Song et~al.(2020)Song, Ding, Dong, and Han}]{song2020distributed}
\bibinfo{author}{H.~Song}, \bibinfo{author}{D.~Ding}, \bibinfo{author}{H.~Dong}, \bibinfo{author}{Q.-L. Han},
\newblock \bibinfo{title}{Distributed maximum correntropy filtering for stochastic nonlinear systems under deception attacks},
\newblock \bibinfo{journal}{IEEE Transactions on Cybernetics}  (\bibinfo{year}{2020}).
\bibitem[{Wang et~al.(2021)Wang, Li, and Zhang}]{wang2021distributed}
\bibinfo{author}{G.~Wang}, \bibinfo{author}{N.~Li}, \bibinfo{author}{Y.~Zhang},
\newblock \bibinfo{title}{Distributed maximum correntropy linear and nonlinear filters for systems with non-gaussian noises},
\newblock \bibinfo{journal}{Signal Processing} \bibinfo{volume}{182} (\bibinfo{year}{2021}) \bibinfo{pages}{107937}.
\bibitem[{Principe(2010)}]{principe2010information}
\bibinfo{author}{J.~C. Principe}, \bibinfo{title}{Information theoretic learning: Renyi's entropy and kernel perspectives}, \bibinfo{publisher}{Springer Science \& Business Media}, \bibinfo{year}{2010}.
\bibitem[{Chen et~al.(2021)Chen, Dang, Gu, Zheng, and Príncipe}]{8937723}
\bibinfo{author}{B.~Chen}, \bibinfo{author}{L.~Dang}, \bibinfo{author}{Y.~Gu}, \bibinfo{author}{N.~Zheng}, \bibinfo{author}{J.~C. Príncipe},
\newblock \bibinfo{title}{Minimum error entropy kalman filter},
\newblock \bibinfo{journal}{IEEE Transactions on Systems, Man, and Cybernetics: Systems} \bibinfo{volume}{51} (\bibinfo{year}{2021}) \bibinfo{pages}{5819--5829}.
\bibitem[{Dang et~al.(2020)Dang, Chen, Wang, Ma, and Ren}]{dang2020robust}
\bibinfo{author}{L.~Dang}, \bibinfo{author}{B.~Chen}, \bibinfo{author}{S.~Wang}, \bibinfo{author}{W.~Ma}, \bibinfo{author}{P.~Ren},
\newblock \bibinfo{title}{Robust power system state estimation with minimum error entropy unscented kalman filter},
\newblock \bibinfo{journal}{IEEE Transactions on Instrumentation and Measurement} \bibinfo{volume}{69} (\bibinfo{year}{2020}) \bibinfo{pages}{8797--8808}.
\bibitem[{Li et~al.(2021)Li, Jing, and Leung}]{li2021robust}
\bibinfo{author}{M.~Li}, \bibinfo{author}{Z.~Jing}, \bibinfo{author}{H.~Leung},
\newblock \bibinfo{title}{Robust minimum error entropy based cubature information filter with non-gaussian measurement noise},
\newblock \bibinfo{journal}{IEEE Signal Processing Letters} \bibinfo{volume}{28} (\bibinfo{year}{2021}) \bibinfo{pages}{349--353}.
\bibitem[{Wang et~al.(2021)Wang, Chen, Yang, Peng, and Feng}]{wang2021numerically}
\bibinfo{author}{G.~Wang}, \bibinfo{author}{B.~Chen}, \bibinfo{author}{X.~Yang}, \bibinfo{author}{B.~Peng}, \bibinfo{author}{Z.~Feng},
\newblock \bibinfo{title}{Numerically stable minimum error entropy kalman filter},
\newblock \bibinfo{journal}{Signal Processing} \bibinfo{volume}{181} (\bibinfo{year}{2021}) \bibinfo{pages}{107914}.
\bibitem[{Erdogmus and Principe(2002)}]{erdogmus2002error}
\bibinfo{author}{D.~Erdogmus}, \bibinfo{author}{J.~C. Principe},
\newblock \bibinfo{title}{An error-entropy minimization algorithm for supervised training of nonlinear adaptive systems},
\newblock \bibinfo{journal}{IEEE Transactions on Signal Processing} \bibinfo{volume}{50} (\bibinfo{year}{2002}) \bibinfo{pages}{1780--1786}.
\bibitem[{Liu et~al.(2006)Liu, Pokharel, and Principe}]{liu2006error}
\bibinfo{author}{W.~Liu}, \bibinfo{author}{P.~Pokharel}, \bibinfo{author}{J.~Principe},
\newblock \bibinfo{title}{Error entropy, correntropy and m-estimation},
\newblock in: \bibinfo{booktitle}{2006 16th IEEE Signal Processing Society Workshop on Machine Learning for Signal Processing}, \bibinfo{organization}{IEEE}, \bibinfo{year}{2006}, pp. \bibinfo{pages}{179--184}.
\bibitem[{Dang et~al.(2022)Dang, Chen, Huang, Zhang, and Zhao}]{9646176}
\bibinfo{author}{L.~Dang}, \bibinfo{author}{B.~Chen}, \bibinfo{author}{Y.~Huang}, \bibinfo{author}{Y.~Zhang}, \bibinfo{author}{H.~Zhao},
\newblock \bibinfo{title}{Cubature kalman filter under minimum error entropy with fiducial points for ins/gps integration},
\newblock \bibinfo{journal}{IEEE/CAA Journal of Automatica Sinica} \bibinfo{volume}{9} (\bibinfo{year}{2022}) \bibinfo{pages}{450--465}. \bibinfo{note}{Doi{: 10.1109/JAS.2021.1004350}}.
\bibitem[{Mohamed and Schwarz(1999)}]{mohamed1999adaptive}
\bibinfo{author}{A.~Mohamed}, \bibinfo{author}{K.~Schwarz},
\newblock \bibinfo{title}{Adaptive kalman filtering for ins/gps},
\newblock \bibinfo{journal}{Journal of geodesy} \bibinfo{volume}{73} (\bibinfo{year}{1999}) \bibinfo{pages}{193--203}.
\bibitem[{Zhao et~al.(2010)Zhao, Wang, Sun, Ding, and Yan}]{zhao2010adaptive}
\bibinfo{author}{L.~Zhao}, \bibinfo{author}{X.-X. Wang}, \bibinfo{author}{M.~Sun}, \bibinfo{author}{J.-C. Ding}, \bibinfo{author}{C.~Yan},
\newblock \bibinfo{title}{Adaptive ukf filtering algorithm based on maximum a posterior estimation and exponential weighting},
\newblock \bibinfo{journal}{Acta Automatica Sinica} \bibinfo{volume}{36} (\bibinfo{year}{2010}) \bibinfo{pages}{1007--1019}.
\bibitem[{Gao et~al.(2015)Gao, Hu, and Zhong}]{gao2015windowing}
\bibinfo{author}{S.~Gao}, \bibinfo{author}{G.~Hu}, \bibinfo{author}{Y.~Zhong},
\newblock \bibinfo{title}{Windowing and random weighting-based adaptive unscented kalman filter},
\newblock \bibinfo{journal}{International Journal of Adaptive Control and Signal Processing} \bibinfo{volume}{29} (\bibinfo{year}{2015}) \bibinfo{pages}{201--223}.
\bibitem[{Van Der~Merwe and Wan(2001)}]{van2001square}
\bibinfo{author}{R.~Van Der~Merwe}, \bibinfo{author}{E.~A. Wan},
\newblock \bibinfo{title}{The square-root unscented kalman filter for state and parameter-estimation},
\newblock in: \bibinfo{booktitle}{2001 IEEE international conference on acoustics, speech, and signal processing. Proceedings (Cat. No. 01CH37221)}, volume~\bibinfo{volume}{6}, \bibinfo{organization}{IEEE}, \bibinfo{year}{2001}, pp. \bibinfo{pages}{3461--3464}.
\bibitem[{Kulikov and Kulikova(2019)}]{kulikov2019numerical}
\bibinfo{author}{G.~Y. Kulikov}, \bibinfo{author}{M.~V. Kulikova},
\newblock \bibinfo{title}{Numerical robustness of extended kalman filtering based state estimation in ill-conditioned continuous-discrete nonlinear stochastic chemical systems},
\newblock \bibinfo{journal}{International Journal of Robust and Nonlinear Control} \bibinfo{volume}{29} (\bibinfo{year}{2019}) \bibinfo{pages}{1377--1395}.
\bibitem[{Chen et~al.(2021)Chen, Xing, Zhao, Du, and Príncipe}]{8793125}
\bibinfo{author}{B.~Chen}, \bibinfo{author}{L.~Xing}, \bibinfo{author}{H.~Zhao}, \bibinfo{author}{S.~Du}, \bibinfo{author}{J.~C. Príncipe},
\newblock \bibinfo{title}{Effects of outliers on the maximum correntropy estimation: A robustness analysis},
\newblock \bibinfo{journal}{IEEE Transactions on Systems, Man, and Cybernetics: Systems} \bibinfo{volume}{51} (\bibinfo{year}{2021}) \bibinfo{pages}{4007--4012}.
\bibitem[{Chen et~al.(2018)Chen, Xing, Xu, Zhao, and Príncipe}]{7795189}
\bibinfo{author}{B.~Chen}, \bibinfo{author}{L.~Xing}, \bibinfo{author}{B.~Xu}, \bibinfo{author}{H.~Zhao}, \bibinfo{author}{J.~C. Príncipe},
\newblock \bibinfo{title}{Insights into the robustness of minimum error entropy estimation},
\newblock \bibinfo{journal}{IEEE Transactions on Neural Networks and Learning Systems} \bibinfo{volume}{29} (\bibinfo{year}{2018}) \bibinfo{pages}{731--737}.
\bibitem[{Chen et~al.(2015)Chen, Wang, Zhao, Zheng, and Principe}]{chen2015convergence}
\bibinfo{author}{B.~Chen}, \bibinfo{author}{J.~Wang}, \bibinfo{author}{H.~Zhao}, \bibinfo{author}{N.~Zheng}, \bibinfo{author}{J.~C. Principe},
\newblock \bibinfo{title}{Convergence of a fixed-point algorithm under maximum correntropy criterion},
\newblock \bibinfo{journal}{IEEE Signal Processing Letters} \bibinfo{volume}{22} (\bibinfo{year}{2015}) \bibinfo{pages}{1723--1727}.
\bibitem[{Zhang et~al.(2015)Zhang, Chen, Liu, Yuan, and Principe}]{zhang2015convergence}
\bibinfo{author}{Y.~Zhang}, \bibinfo{author}{B.~Chen}, \bibinfo{author}{X.~Liu}, \bibinfo{author}{Z.~Yuan}, \bibinfo{author}{J.~C. Principe},
\newblock \bibinfo{title}{Convergence of a fixed-point minimum error entropy algorithm},
\newblock \bibinfo{journal}{Entropy} \bibinfo{volume}{17} (\bibinfo{year}{2015}) \bibinfo{pages}{5549--5560}.
\bibitem[{Kulikov and Kulikova(2021)}]{kulikov2021ito}
\bibinfo{author}{G.~Y. Kulikov}, \bibinfo{author}{M.~V. Kulikova},
\newblock \bibinfo{title}{Ito\^{}-taylor-based square-root unscented kalman filtering methods for state estimation in nonlinear continuous-discrete stochastic systems},
\newblock \bibinfo{journal}{European Journal of Control} \bibinfo{volume}{58} (\bibinfo{year}{2021}) \bibinfo{pages}{101--113}.

\end{thebibliography}





\end{document}